\documentclass[
    aip,
    jcp,
    reprint,
    superscriptaddress,
    amsmath,
    amssymb
]{revtex4-1}

\pdfoutput=1

\usepackage{graphicx}
\usepackage{dcolumn}
\usepackage{bm}
\usepackage{amsthm}
\usepackage{float}
\usepackage{booktabs}
\usepackage{siunitx}
\usepackage{multirow}

\usepackage[english]{babel}
\renewcommand{\selectlanguage}[1]{}

\usepackage[
    hyperindex,
    breaklinks,
    hidelinks,
    colorlinks,
    citecolor=black,
    linkcolor=black,
    urlcolor=black
]{hyperref}

\begin{document}

\title{Integrated Alchemical and Conformational Enhanced Sampling for Solvation Free Energy Calculations}

\author{Gabriela B. Correa}
\affiliation{Department of Chemistry, New York University, New York City, New York 10003, United States}

\author{Charlles R. A. Abreu}
\affiliation{Department of Chemical Engineering, Massachusetts Institute of Technology, Cambridge, Massachusetts 02139, United States}

\author{Nisanth N. Nair}
\affiliation{Department of Chemistry, Indian Institute of Technology Kanpur, Kanpur 208016, India}

\author{Mark E. Tuckerman}
\email{mark.tuckerman@nyu.edu}
\affiliation{Department of Chemistry, New York University, New York City, New York 10003, United States}
\affiliation{Simons Center for Computational Physical Chemistry, New York University, New York City, New York 10003, United States}
\affiliation{Courant Institute of Mathematical Sciences, New York University, New York City, New York 10012, United States}

\begin{abstract}
Accurate solvation free energies from molecular dynamics simulations require efficient sampling of coupled slow variables, including solvent coordinates, solute conformational modes, and the alchemical coordinate $\lambda$. 
Here, we develop a $\lambda$-dynamics framework that combines mass scaling, on-the-fly probability enhanced sampling (OPES), and driven adiabatic free energy dynamics (d-AFED) to address these sampling challenges within a unified protocol. 
For rigid organic solutes, Hamiltonian replica exchange with mass scaling is first used to quantify the effect of octanol solvent relaxation. Reducing all octanol atomic masses by a factor of ten accelerates convergence by more than fivefold while preserving equilibrium solvation free energies. 
These calculations then provide reference benchmarks for $\lambda$-OPES, a dual-bias $\lambda$-dynamics strategy that combines the ``standard'' and ``explore'' variants of OPES to promote transitions along the alchemical coordinate. This approach reaches convergence on timescales comparable to replica exchange, but without predefined $\lambda$ windows or multiple parallel simulations. 
For flexible $N$-acetyl amino-acid amide solutes, $\lambda$-OPES is coupled with d-AFED on selected backbone and side-chain dihedrals to enable simultaneous alchemical and conformational enhanced sampling. This combined strategy improves agreement with experimental octanol-water partition coefficients and reduces the mean absolute error from 0.75 log units with $\lambda$-OPES alone to 0.30 log units with $\lambda$-OPES-d-AFED. 
Overall, this work establishes an integrated enhanced sampling protocol for solvation free energy calculations across rigid organic solutes and flexible peptide-like solutes, and provides a foundation for the application of alchemical free energy methods to larger and more conformationally complex systems.
\end{abstract}

\maketitle

\section{Introduction}

Solvation free energies provide a thermodynamic route to hydration and transfer processes across diverse solutes and solvents. 
For example, the octanol-water partition coefficient, a widely adopted measure of molecular lipophilicity\cite{Arnott_2012, Bayliss_2016, Johnson_2018} with direct relevance to drug discovery,\cite{Sala_2021,Wang_2022,Fetse_2023,Zheng_2025} can be obtained from the difference between solvation free energies in water and octanol. 
This connection has made solvation a central application of free energy methods in molecular dynamics (MD) simulations.\cite{Bannan_2016, Espinosa_2018, Nedyalkova_2019, Fan_2020, Sabatino_2021}
By combining thermodynamic rigor with molecular resolution, MD-based approaches are powerful, but also sensitive to sampling limitations\cite{Kang_2026}. 
Reliable predictions require methods that can efficiently explore the coupled slow degrees of freedom involved in solvation.\cite{Mey_2020, Henin_2022}

In MD simulations, solvation free energies are commonly computed using alchemical methods\cite{Hansen_2014, Klimovich_2015}, in which a continuous parameter $\lambda$ scales the interactions between the solute and its environment.
Along this pathway, $\lambda=0$ corresponds to a non-interacting solute and $\lambda=1$ to a fully interacting solute in solution, with the solvation free energy obtained from the reversible work associated with the transformation.
The accuracy and efficiency of the calculation depend critically on how the $\lambda$ domain is sampled.

Conventional approaches\cite{Naden_2014, Naden_2015, Abreu_2020a, Correa_2022} use multiple fixed-$\lambda$ simulations, often requiring many intermediate states to maintain phase-space overlap along the transformation pathway.
Hamiltonian replica exchange (HREX)\cite{Meng_2011, Chodera_2011} improves sampling by allowing configurations to exchange between neighboring $\lambda$ states, which promotes mixing and helps replicas escape local minima. However, it still requires careful spacing along $\lambda$, multiple coordinated simulations, and efficient communication between neighboring replicas.
An alternative strategy is to make $\lambda$ itself a dynamical variable. In $\lambda$-dynamics,\cite{Kong_1996,Abrams_2006,Knight_2009} the alchemical coordinate is propagated together with the atomic coordinates, allowing sampling within a single simulation. This formulation removes the need to prescribe a discrete set of states, but introduces a different problem: the trajectory must repeatedly cross barriers along $\lambda$ and visit the physical end states often enough to reconstruct the free energy profile.

Several enhanced sampling strategies have been developed to address this issue\cite{Bieler_2014, Robo_2023}.
$\lambda$-AFED\cite{Abrams_2006} accelerates the alchemical coordinate by coupling it to high-temperature auxiliary dynamics, requiring large fictitious masses, temperatures $>10000$ K, and careful thermostatting.
$\lambda$-metadynamics\cite{Wu_2011} reconstructs the free energy profile with history-dependent Gaussian biases, and its performance depends on several parameters.
Adaptive landscape flattening in multisite $\lambda$-dynamics\cite{Hayes_2017, Hayes_2021, Hayes_2024} facilitates transitions between states by applying system-specific bias potentials learned from preliminary simulations.
More recently, $\lambda$-ABF\cite{Louis_2024} has promoted nearly uniform exploration using adaptive biasing forces, with smooth initial scaling used to control the large variance of early force estimates along $\lambda$.
Together, these methods demonstrate that biasing $\lambda$ can substantially improve convergence, while introducing different compromises in parameter selection, prior knowledge, and numerical stability.

On-the-fly probability enhanced sampling (OPES)\cite{Invernizzi_2020} offers a promising adaptive bias framework for $\lambda$-dynamics.
Inspired by well-tempered metadynamics, OPES constructs a bias potential from a real-time probability estimate and drives the system toward a prescribed target distribution using a small number of physically motivated parameters.
Its two main variants serve complementary purposes:\cite{Invernizzi_2022} OPES-Standard accelerates free energy convergence, and OPES-Explore promotes rapid barrier crossing and access to metastable states.
OPES-Explore was recently introduced as an auxiliary bias within the $\lambda$-ABF-OPES method,\cite{Ansari_2025, Ansari_2026} yielding a ninefold acceleration in absolute binding free energy calculations.
Whether OPES can serve as the primary bias acting directly on $\lambda$, and which variant or combination provides the most robust strategy for solvation free energy calculations, remain open questions.

As solute-solvent interactions are controlled by $\lambda$, the surrounding liquid must reorganize, and slow solvent dynamics can delay convergence even when transitions along $\lambda$ occur frequently.
This issue is particularly relevant for organic solvents such as octanol, which is relatively large, viscous, and structurally heterogeneous,\cite{Hoffmann_2024} leading to slow collective rearrangements around the solute.
Because solvent molecules constitute most of the simulated system, strategies to accelerate solvent relaxation should be inexpensive and general, without relying on high-dimensional solvent collective variables defined per molecule.
Mass scaling and mass repartitioning provide such a route by modifying solvent inertia relative to the solute.
In classical statistical mechanics, equilibrium configurational properties are independent of atomic masses, so solvent masses can be modified to accelerate dynamics without changing the target thermodynamic ensemble.
Although these approaches have improved sampling efficiency in aqueous\cite{Jimenez_2024} and biomolecular systems,\cite{Lin_2010, Hopkins_2015, Pang_2015, Bowers_2026} their impact on solvation free energy calculations in organic solvents has not been assessed.

Slow intramolecular motions of the solute introduce an additional sampling requirement.
This is especially important for peptide-like solutes, where backbone and side-chain dihedrals, intramolecular hydrogen bonding, and solvent exposure are strongly coupled.\cite{Gopal_2017, Yadav_2019}
Previous simulations of blocked amino acids,\cite{Konig_2013, Khoury_2014, Hajari_2015, Schauperl_2016} oligoglycines,\cite{Hu_2010, Drake_2016} cyclic peptides,\cite{Tomar_2013, Lou_2025} and polypeptides\cite{Foumthuim_2023, Foumthuim_2024} have provided valuable insight into solvation, but most relied on conventional MD without targeted acceleration of relevant torsional modes.
Only a few studies have explicitly coupled alchemical calculations with torsion sampling,\cite{Cuendet_2012, Hsu_2023, Verma_2024, Zhou_2026, Verma_2026} highlighting the need for strategies that can accelerate conformational relaxation within solvation free energy calculations.

Torsional modes can, in principle, be accelerated using adaptive biasing methods such as metadynamics, ABF, and OPES. 
However, these approaches are most effective in low-dimensional spaces\cite{Pfaendtner_2015, Awasthi_2017}, typically involving only one to three collective variables. Applying a single adaptive bias across the joint space of multiple torsions would require constructing a high-dimensional free energy or mean force surface on the fly, whose computational demand increases with molecular flexibility.
A more natural and scalable strategy is provided by driven adiabatic free energy dynamics (d-AFED)\cite{Rosso_2002}, which can accelerate multiple collective variables, such as torsion angles, simultaneously without the need to build an explicit multidimensional bias. This advantage has already been demonstrated in peptide conformational sampling\cite{Bajpai_2023, Hong_2024, Bajpai_2025}. By coupling selected dihedral angles to auxiliary variables with appropriately chosen fictitious masses, force constants, and temperatures around 1000-3000 K, d-AFED promotes rapid barrier crossing and maintains adiabatic separation from the physical system.
Therefore, a promising division of labor is to use OPES to drive exploration along $\lambda$ and reconstruct the solvation free energy profile, and to use d-AFED to accelerate torsional modes and promote conformational relaxation, rather than to reconstruct the torsional free energy landscape in detail.

In this work, we target the three coupled sources of slow convergence in solvation free energy calculations discussed above: alchemical sampling, solvent relaxation, and solute conformational relaxation. 
To address them within a unified $\lambda$-dynamics framework, we combine solvent mass scaling, OPES biasing along the alchemical coordinate, and d-AFED acceleration of selected torsional modes, and evaluate the resulting protocol in water and octanol for systems ranging from rigid organic solutes to flexible peptide-like solutes. 
We first use HREX simulations to assess octanol mass scaling and quantify its effect on convergence and equilibrium free energies for rigid organic solutes. 
We then develop and benchmark $\lambda$-OPES as an adaptive-bias $\lambda$-dynamics approach, comparing OPES-Standard, OPES-Explore, and a dual-bias combination against HREX reference calculations. 
Finally, we couple the best-performing $\lambda$-OPES protocol to d-AFED on selected dihedral angles in flexible $N$-acetyl amino-acid amide solutes, yielding a $\lambda$-OPES-d-AFED strategy for simultaneous alchemical and conformational enhanced sampling. 
All calculations are performed within the linear basis function (LBF) formulation for solute-solvent interactions developed in our previous work,\cite{Correa_2022} and the resulting free energies are validated against available experimental octanol-water partition coefficient data.\cite{Hansch_1995, Sangster_1989, Leo_1971, Fauchere_1983}

The remainder of this article is organized as follows. 
Section~2 describes the simulation methodology, including the $\lambda$-dynamics formulation, solute-solvent interactions, and the OPES and d-AFED enhanced sampling protocols. 
Section~3 presents the results, beginning with the effect of octanol mass scaling on convergence, followed by the development of OPES-biased $\lambda$-dynamics and its extension to conformationally flexible solutes through $\lambda$-OPES-d-AFED. 
Section~4 summarizes the main conclusions.

\section{Methodology}

\subsection{$\lambda$-Dynamics Formulation}
\label{methods_lambda_dynamics}

We employ two extended-dynamics formulations: $\lambda$-OPES and $\lambda$-OPES-d-AFED. In both cases, the alchemical coordinate $\lambda$ is propagated as a dynamical variable. In $\lambda$-OPES-d-AFED, selected solute torsions $\bm{\xi}=\{\xi_k\}$ are additionally coupled to auxiliary variables $\mathbf{s}=\{s_k\}$ to enhance conformational sampling. The extended Hamiltonian is written as\cite{Chen_2012}
\begin{equation}
\label{eq:extended_hamiltonian}
\begin{aligned}
\mathcal{H} = 
&\sum_i \frac{\mathbf{p}_i^2}{2m_i} 
+ \frac{p_\lambda^2}{2m_\lambda} 
+ \sum_k \frac{p_{s_k}^2}{2m_{s_k}} \\
&+ U_\lambda(\mathbf{r},\lambda) 
+ U_{\mathrm{OPES}}(\lambda,t) 
+ U_{\mathrm{d{\text -}AFED}}(\bm{\xi}(\mathbf{r}),\mathbf{s}) .
\end{aligned}
\end{equation}
The first line contains the kinetic energies of the atomic, alchemical, and d-AFED auxiliary degrees of freedom, respectively, whereas the second line contains the corresponding potential energy terms. Here, $\mathbf{r} \equiv \mathbf{r}_1,...,\mathbf{r}_N$ denotes the full set of atomic coordinates; $\mathbf{p}_i$ and $m_i$ are the momenta and mass of atom $i$; $(p_\lambda,m_\lambda)$ and $(p_{s_k},m_{s_k})$ are the momenta and masses assigned to $\lambda$ and $s_k$, respectively. The term $U_\lambda$ defines the alchemical potential energy pathway, including the solute and solvent contributions and the $\lambda$-dependent solute-solvent interactions. The OPES bias potential, $U_{\mathrm{OPES}}$, acts on $\lambda$, and $U_{\mathrm{d{\text -}AFED}}$ couples the auxiliary variables $\mathbf{s}$ to the selected torsional collective variables. In $\lambda$-OPES simulations, the $\mathbf{s}$-dependent kinetic energy and $U_{\mathrm{d{\text -}AFED}}$ terms are omitted. Detailed expressions for $U_\lambda$, $U_{\mathrm{OPES}}$, and $U_{\mathrm{d{\text -}AFED}}$ are given in Sections~\ref{methods_lbf}, \ref{methods_opes}, and \ref{methods_afed}, respectively.

Each set of degrees of freedom is coupled to its own thermostat.
The atomic and alchemical coordinates are maintained at the target temperature $T$, and each d-AFED auxiliary variable is thermostatted at $T_{s_k}$.
The masses of the extended variables, $m_\lambda$ and $m_{s_k}$, are chosen to maintain approximate adiabatic separation\cite{Rosso_2002} from the physical degrees of freedom. In this regime, the physical coordinates relax rapidly relative to the evolution of the extended variables, so that the system remains close to equilibrium for each instantaneous value of $\lambda$ and $\mathbf{s}$. 
Numerically, the physical and extended variables are advanced with separate integrators using the same time step; after each step, their updated values are used to evaluate the coupling forces for the next step.

Solvation free energies are obtained by thermodynamic integration of the physical mean force along $\lambda$,\cite{BitettiPutzer_2003}
\begin{equation}
\label{eq:deltaG_mean_force}
\Delta G =
\int_{0}^{1}
\left\langle
\frac{\partial U_\lambda}{\partial \lambda}
\right\rangle_{\lambda}
 d\lambda ,
\end{equation}
where $\langle \cdots \rangle_{\lambda}$ denotes an ensemble average conditioned on a given value of $\lambda$. In practice, $\partial U_\lambda/\partial\lambda$ is accumulated in bins along $\lambda$, and the resulting mean force profile is integrated numerically. Under adiabatic separation, this conditional mean force is unaffected by changes in the marginal sampling of $\lambda$, including those introduced by the auxiliary temperature and the OPES bias.\cite{Chen_2012,Cuendet_2014} When d-AFED is applied to the auxiliary torsional variables, the sampled configurations can be reweighted according to the procedure of \citet{Cuendet_2012} to recover the corresponding physical torsional populations. Applying this reweighting produced only negligible changes in the solvation free energies. Therefore, the solvation free energies reported here are obtained directly from the mean force accumulated during the d-AFED simulations, without torsional reweighting.

To avoid boundary artifacts at the end states $\lambda=0$ and $\lambda=1$, we use a mirror-periodic representation of the alchemical coordinate, following \citet{Wu_2011}. A periodic variable $\theta$ evolves in the interval $[0,2]$ and is mapped onto the physical interval $\lambda\in[0,1]$ as
\begin{equation}
\label{eq:theta_mapping}
\lambda = 
\begin{cases}
\theta, & 0 \leq \theta \leq 1, \\
2 - \theta, & 1 < \theta \leq 2 .
\end{cases}
\end{equation}
When $\theta$ lies in $(1,2]$, the system follows the reverse alchemical path, and the generalized force changes sign accordingly. Values of $\theta$ outside $[0,2]$ are mapped back into the interval using periodic boundary conditions. This construction allows the OPES bias to populate the free energy basins near both end states.

\subsection{Linear Basis Function}
\label{methods_lbf}

We adopt the coupling route\cite{Klimovich_2015}, in which only solute-solvent interactions are alchemically transformed. Following the LBF formulation\cite{Naden_2014, Naden_2015} developed in our previous work,\cite{Correa_2022} the alchemical potential is written as
\begin{equation}
\begin{aligned}
    U_{\lambda}(\mathbf{r},\lambda) &=
    U_{\text{solv}}(\mathbf{r}) +
    U_{\text{solu}}(\mathbf{r})  \\
    &\quad +
    h_A(\lambda)\, U_{\text{solu-solv}}(\mathbf{r}) +
    h_B(\lambda)\, U_{\mathrm{softLJ}}(\mathbf{r}) .
\end{aligned}
\label{eq:lbf_potential}
\end{equation}
The terms $U_{\text{solv}}$ and $U_{\text{solu}}$ contain all bonded interactions and solvent-solvent or solute-solute nonbonded interactions, and remain unchanged along the alchemical transformation.
The term $U_{\text{solu-solv}}$ represents the full solute-solvent nonbonded interaction, including Lennard-Jones and Coulomb contributions. The additional basis potential $U_{\mathrm{softLJ}}$ is a softened Lennard-Jones interaction used to avoid endpoint singularities when the solute-solvent interactions are weakly coupled. In this formulation, the $\lambda$ dependence enters only through the switching functions $h_A(\lambda)$ and $h_B(\lambda)$, making the potential linear in the basis functions and avoiding the nonlinear dependence of conventional soft-core potentials\cite{Beutler_1994}.

The softened Lennard-Jones basis is defined as\cite{Correa_2022}
\begin{equation}
    U_{\mathrm{softLJ}}(\mathbf{r}) =
    \sum_{i \in \mathrm{solu}} \sum_{j \in \mathrm{solv}}
    \epsilon_{ij}\, u_{\mathrm{cap}}(r_{ij}/\sigma_{ij}),
\label{eq:soft_lj_total}
\end{equation}
\begin{equation}
    u_{\mathrm{cap}}(x) =
    \begin{cases}
        -868x^6 + 2188.8x^5 - 1440x^4 + 119.2, & x < 1, \\[6pt]
        4x^{-12} - 4x^{-6}, & x \geq 1 .
    \end{cases}
\label{eq:soft_lj_reduced}
\end{equation}
For $r_{ij}\geq\sigma_{ij}$, this potential is identical to the standard Lennard-Jones interaction. At shorter distances, the divergent repulsive wall is replaced by a finite polynomial cap, which approaches $119.2\epsilon_{ij}$ as $r_{ij}\to 0$. The cap is smoothly joined to the Lennard-Jones potential at $r_{ij}=\sigma_{ij}$, avoiding discontinuities in the potential and its derivatives.

The switching functions define a continuous pathway from the non-interacting to the fully interacting solute. The softened Lennard-Jones basis is introduced at small $\lambda$ to avoid endpoint singularities and is subsequently removed as the full solute-solvent Lennard-Jones and Coulomb interactions are activated. To reduce the risk of numerical instabilities, the full solute-solvent interaction starts contributing only after $\lambda=0.35$\cite{Correa_2022}
\begin{equation}
    h_A(\lambda) = S_{32}\left( \frac{\lambda - 0.35}{0.65} \right),
\label{eq:hA}
\end{equation}
\begin{equation}
    h_B(\lambda) = S_{22}(\lambda) - S_{32}\left( \frac{\lambda - 0.35}{0.65} \right),
\label{eq:hB}
\end{equation}
where $S_{mn}(x)$ is a smoothstep function with $m$ and $n$ vanishing derivatives at $x=0$ and $x=1$, respectively. The specific functions used are
\begin{equation}
    S_{22}(x) = 6x^5 - 15x^4 + 10x^3,
    \qquad 0 \leq x \leq 1,
\label{eq:S22}
\end{equation}
\begin{equation}
    S_{32}(x) =
    \begin{cases}
        0, & x \leq 0, \\[4pt]
        10x^6 - 24x^5 + 15x^4, & 0 < x \leq 1 .
    \end{cases}
\label{eq:S32}
\end{equation}
With this choice, $h_A=0$ for $\lambda\leq0.35$, so only the softened Lennard-Jones basis contributes during the early part of the coupling pathway. For $\lambda>0.35$, $h_A$ smoothly turns on the full solute-solvent interaction while $h_B$ removes the softened basis. At $\lambda=1$, $h_A=1$ and $h_B=0$, recovering the fully interacting physical potential.

\subsection{On-the-fly Probability Enhanced Sampling}
\label{methods_opes}

OPES is used to construct an adaptive bias from a real-time estimate of the probability distribution along $\lambda$. At simulation time $t$, the bias potential is written as\cite{Invernizzi_2020}
\begin{equation}
\label{eq:opes_bias_potential}
U_{\text{OPES}}(\lambda,t) =
\frac{\nu}{\beta}
\ln \left(
\frac{\hat{\rho}(\lambda,t)}{\overline{\rho}(t)}
+ \epsilon
\right),
\end{equation}
where $\hat{\rho}(\lambda,t)$ is the kernel density estimate accumulated up to time $t$, $\overline{\rho}(t)$ is the average of $\hat{\rho}(\lambda,t)$ over the centers of the compressed kernels, $\epsilon$ is a regularization constant that sets a lower bound for the argument of the logarithm, $\nu$ is a coefficient that depends on the bias factor $\gamma$, and $\beta=(k_{\mathrm B}T)^{-1}$, with $k_{\mathrm B}$ denoting the Boltzmann constant.

The OPES barrier parameter, $\Delta E$, approximately represents the free energy barrier to be crossed during the simulation. It sets the bias factor $\gamma$ and the regularization constant $\epsilon$ through
\begin{equation}
\gamma=\beta\Delta E,
\end{equation}
\begin{equation}
\epsilon = \exp\left(-\frac{\beta \Delta E}{\nu}\right).
\end{equation}
Together, the normalization by $\overline{\rho}(t)$ and the regularization by $\epsilon$ improve the numerical stability of the adaptive bias. The normalization reduces sensitivity to changes in the estimated probability distribution as new regions of $\lambda$ space are explored.

Two OPES variants are considered: OPES-Standard\cite{Invernizzi_2020} and OPES-Explore.\cite{Invernizzi_2022} In OPES-Standard, the kernels are reweighted so that $\hat{\rho}(\lambda,t)$ converges to the unbiased probability distribution, $\rho^\ast(\lambda)$, generally yielding a smoother and more stable bias. In OPES-Explore, all kernels are assigned equal weights, so that $\hat{\rho}(\lambda,t)$ converges to the sampled distribution, $\rho(\lambda)$, leading to more aggressive exploration of poorly sampled regions. With $\nu=1-\gamma^{-1}$ for OPES-Standard and $\nu=\gamma-1$ for OPES-Explore, both variants target a well-tempered distribution proportional to $[\rho^\ast(\lambda)]^{1/\gamma}$.

Based on these complementary modes of behavior, we examine both single-bias and dual-bias protocols. In the single-bias cases, either OPES-Standard or OPES-Explore is applied independently. In the dual-bias protocol, the two variants are constructed and updated separately but applied concurrently, so that the total bias is
\begin{equation}
U_{\text{OPES}}(\lambda,t)
=
U_{\text{OPES-Stand}}(\lambda,t)
+
U_{\text{OPES-Expl}}(\lambda,t).
\end{equation}
The resulting trajectory is generated by summing both contributions and is used to update each of them. An explicit characterization of the distribution produced by the combined adaptive bias is not required, because the free energy profile is reconstructed from the conditional mean generalized force. This protocol is designed to retain the smoother reweighting-based convergence of OPES-Standard while benefiting from the stronger exploratory character of OPES-Explore.

As with metadynamics, the practical application of OPES is limited by the dimensionality of the collective-variable space, since estimating and biasing a multidimensional probability distribution becomes increasingly difficult as the number of variables grows. For this reason, OPES is applied only to the one-dimensional alchemical coordinate $\lambda$, rather than to all auxiliary variables simultaneously.

\subsection{Adiabatic Free Energy Dynamics}
\label{methods_afed}

The d-AFED method is used to enhance conformational sampling of backbone and side-chain dihedral angles in the $N$-acetyl amino-acid amide solutes. Each torsional collective variable $\xi_k$ is coupled to an auxiliary dynamical variable $s_k$ through a harmonic potential,\cite{Chen_2012, Bajpai_2023}
\begin{equation}
U_{\mathrm{d{\text -}AFED}}(\boldsymbol{\xi}(\mathbf r), \mathbf s)
=
\sum_k \frac{\kappa_k}{2}
\left[\xi_k(\mathbf r) - s_k\right]^2,
\end{equation}
where $\kappa_k$ is the coupling force constant, and the difference $\xi_k-s_k$ is evaluated as a periodic angular displacement.

The auxiliary variables are propagated at temperatures higher than the target simulation temperature, thereby promoting torsional barrier crossing and accelerating conformational relaxation. To avoid spurious heat transfer from the hot auxiliary variables to the physical system, the d-AFED parameters are chosen to approximately preserve adiabatic separation between the auxiliary and physical degrees of freedom. Following previous d-AFED applications,\cite{Cuendet_2014, Bajpai_2023} the auxiliary masses $m_{s_k}$ are selected to be about 300 times larger than the typical effective masses of the corresponding torsional collective variables. 
This is based on the formula derived by Cuendet et al.~\cite{Cuendet_2014},
\begin{equation}
    m_{s_k} = c\left[\sum_{i} m_i^{-1}\left\|{\partial \xi_k \over \partial  {\bf r}_i}\right\|^2\right]^{-1},
    \label{eq:aux_masses}
\end{equation}
where $c$ is a constant used to ensure adiabatic decoupling.
The coupling constants $\kappa_k$ are then estimated by targeting an approximate oscillation period of 40~fs for the coupled $\xi_k$--$s_k$ motion, providing a practical balance between responsive coupling and stable numerical integration with the MD time step.

\subsection{Octanol-Water Partition Coefficient}

The octanol-water partition coefficient is computed from the transfer free energy between the two liquid phases,\cite{Espinosa_2018}
\begin{equation}
\log P_{\mathrm{oct/water}} =
\frac{\Delta G_{\mathrm{water}} - \Delta G_{\mathrm{oct}}}
{k_{\mathrm B} T\ln 10},
\end{equation}
where $\Delta G_{\mathrm{water}}$ and $\Delta G_{\mathrm{oct}}$ are the solvation free energies in water and octanol, respectively. With this sign convention, positive values of $\log P_{\mathrm{oct/water}}$ indicate preferential partitioning into octanol, and negative values indicate preferential partitioning into water.

Calculated partition coefficients are compared with reference data from two sources. The benchmark study of \citet{Bannan_2016} reports simulation results for neutral organic compounds together with experimental values compiled from the literature. Experimental values for capped amino-acid derivatives are taken from \citet{Fauchere_1983}. All calculated solvation free energies are obtained from independent single-solute simulations in each solvent and therefore correspond to the infinite-dilution limit. Water and octanol are modeled as separate phases, i.e., their mutual miscibility is neglected.

\subsection{Simulation Details}

\begin{figure}[!htbp]
    \centering
    \includegraphics[scale=0.55]{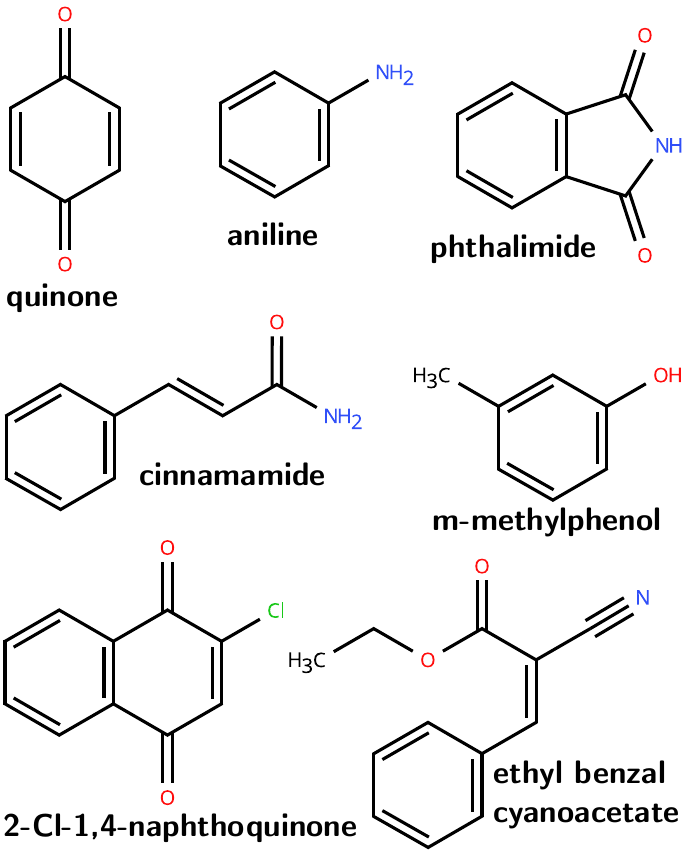}
    \caption{Chemical structures of the rigid organic solutes.}
    \label{fig:organic}
\end{figure}

\begin{figure*}[!htbp]
    \centering
    \includegraphics[scale=0.55]{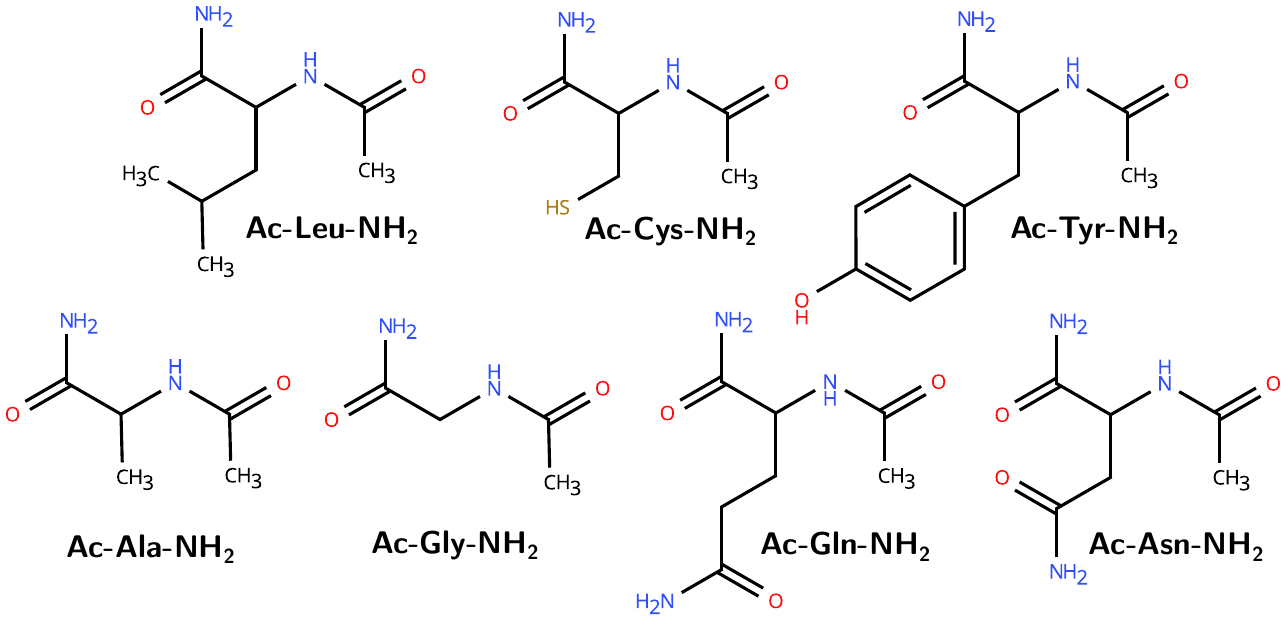}
    \caption{Chemical structures of the flexible peptide-like solutes.}
    \label{fig:peptides}
\end{figure*}

The simulation systems comprise two solute groups: a set of seven rigid organic molecules and another set of seven flexible peptide-like $N$-acetyl amino acid amides, shown in Figures~\ref{fig:organic} and \ref{fig:peptides}, respectively. For the rigid organic set, the solutes are modeled with GAFF\cite{Wang_2004} and AM1-BCC charges,\cite{Jakalian_2000} and are simulated in TIP3P water\cite{Jorgensen_1983} and GAFF octanol. For the peptide-like set, the solutes are modeled with AMBER14SB,\cite{Maier_2015} and the solvent phases are OPC water\cite{Izadi_2014} and GAFF2\cite{ambertools_2023} octanol. Initial configurations are generated with PACKMOL,\cite{Martinez_2009} and topology files are prepared with LEaP. Each simulation box contains one solute molecule and either 500 water molecules or 80 octanol molecules.

All MD simulations are performed with OpenMM version 8.4.0~\cite{Eastman_2024} in the isothermal-isobaric ensemble at 298~K and 1~atm. Pressure is controlled with a Monte Carlo barostat~\cite{Aqvist_2004}, and volume changes are attempted every 50~fs. The atomic degrees of freedom are thermostatted using Langevin dynamics with a friction coefficient of 10~ps$^{-1}$ and propagated with a leapfrog integrator\cite{Zhang_2019}. The extended variables are coupled to a Regulated Nosé--Hoover--Langevin thermostat~\cite{Abreu_2021} with a time constant of 40~fs, a friction coefficient of 25~ps$^{-1}$, and a regulation parameter of 1. 

Unless otherwise stated, the $\lambda$ mass is set to 1~Da\,nm$^2$, and $\lambda$ is maintained at 298~K. In d-AFED simulations, the auxiliary torsional variables are maintained at 1000~K with masses of $50~\mathrm{Da\,nm^2\,rad^{-2}}$ and harmonic coupling constants of $1000~\mathrm{kJ\,mol^{-1}\,rad^{-2}}$.
Long-range electrostatics are treated using the Particle Mesh Ewald method.\cite{Darden_1993,Essmann_1995} Nonbonded interactions are truncated at 1.2~nm, with a switching function starting at 1.1~nm. Bonds involving hydrogen atoms are constrained, and water molecules are kept rigid. A time step of 1~fs is used, and configurations are saved every 200~fs. Production simulations are initiated from configurations equilibrated for 1~ns. For each system and simulation protocol, three independent runs are performed, and the reported statistical uncertainties correspond to the standard error of the mean across these runs.

HREX simulations are performed with \texttt{openmmtools} and used as reference calculations. Each HREX simulation uses 21 evenly spaced $\lambda$ states from 0 to 1. Exchanges are attempted every 1~ps, and each replica is propagated for up to 10~ns, yielding a total sampling time of 210~ns per system.
The $\lambda$-dynamics protocols are simulated for up to 50~ns. The OPES bias potentials are constructed along the transformed alchemical coordinate $\theta$, using a Gaussian width of 0.10, a grid of 101 points, and a kernel deposition stride of 400~fs. For the rigid organic solutes, single-bias OPES-Standard and OPES-Explore simulations use a barrier of 30~kcal~mol$^{-1}$. In the dual-bias protocol, the OPES-Standard and OPES-Explore barriers are set to 30 and 5~kcal~mol$^{-1}$, respectively. For the peptide-like solutes, the corresponding OPES-Standard and OPES-Explore barriers are set to 50 and 5~kcal~mol$^ {- 1} $, respectively.

All extended phase-space simulations are performed with OpenXPS, available at \url{github.com/craabreu/openxps}. The alchemical decomposition of nonbonded interactions is available at \url{github.com/craabreu/openmm-nonbonded-slicing}. Example scripts for the $\lambda$-dynamics protocols are provided at \url{github.com/gabriela-correa/lambda-OPES-d-AFED}.

\section{Results and Discussion}

\subsection{Effect of Octanol Mass Scaling on Convergence}

\begin{figure}[!t]
    \centering
    \includegraphics[scale=0.85]{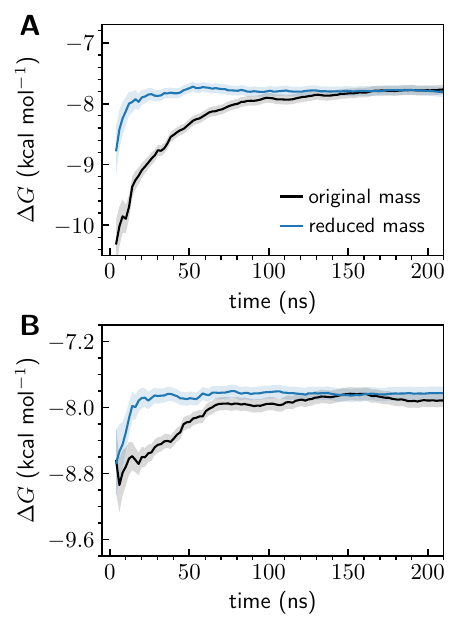}
    \caption{Convergence of solvation free energies for (A) quinone and (B) \textit{m}-methylphenol in octanol obtained from HREX simulations using the original octanol masses and 10-fold reduced octanol masses. The simulation time corresponds to the cumulative sampling time summed over all replicas. The shaded regions represent the corresponding uncertainties.}
    \label{fig:scaling}
\end{figure}

Liquid octanol relaxes more slowly than water, which can delay solvent reorganization around the solute and hinder the convergence of solvation free energy calculations. We examine whether accelerating octanol dynamics through mass scaling improves the convergence of HREX calculations in this solvent. Seven rigid organic solutes are simulated using either the real octanol atomic masses or a modified model in which all atomic masses are reduced by a factor of 10. The solute masses are kept unchanged in all simulations. Convergence profiles for quinone and \textit{m}-methylphenol are presented in Figures~\ref{fig:scaling} A and B, respectively, with the profiles for all solutes reported in Figure~S1.

Reducing the octanol masses substantially accelerates convergence across the full set of solutes. With the original mass assignment, stable solvation free energy estimates are typically obtained only after approximately 100--150 ns of accumulated simulation time. In contrast, the reduced mass octanol model reaches stable estimates within about 20 ns, corresponding to a more than fivefold reduction in the simulation time required for convergence.

The final solvation free energies obtained with the original and reduced octanol masses remain in close agreement, confirming that mass scaling does not measurably alter the equilibrium free energy estimates. The benefit of the reduced mass model is dynamical rather than thermodynamic: decreasing the octanol masses accelerates solvent relaxation and improves configurational sampling without changing the target equilibrium distribution. Larger mass reductions were also tested, but scaling beyond a factor of 10 led to numerical instabilities under the integration conditions and would require a time step smaller than 1 fs.

All simulations in this section use GAFF for octanol, but similar qualitative behavior is expected for other atomistic octanol models because the acceleration arises from the mass dependence of the dynamics rather than from force-field-specific energetic terms. Unless otherwise noted, all subsequent octanol simulations, including those involving peptides and different enhanced sampling strategies, are performed using octanol atomic masses reduced by a factor of 10.

\subsection{Alchemical Sampling with $\lambda$-Dynamics}

We next examine $\lambda$-dynamics\cite{Kong_1996,Abrams_2006,Knight_2009} as an alternative approach for alchemical sampling. Whereas HREX samples a discrete set of fixed alchemical states, $\lambda$-dynamics treats the alchemical coordinate as a continuous dynamical variable that evolves together with the atomic coordinates. This formulation makes the biasing strategy central to promoting efficient transitions between the physical end states while maintaining accurate free energy estimates.

We compare HREX with two OPES-based $\lambda$-dynamics protocols using the same set of rigid organic solutes: $\lambda$-OPES-Explore, which uses only the exploratory OPES-Explore bias, and $\lambda$-OPES, which combines OPES-Standard and OPES-Explore. A protocol using only OPES-Standard was also tested but omitted from further analysis because some systems exhibited too few transitions between $\lambda=0$ and $\lambda=1$. 

\begin{figure}[!t]
    \centering
    \includegraphics[scale=0.85]{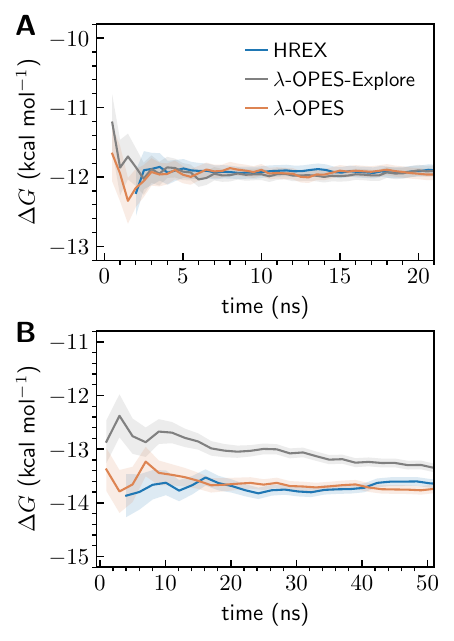}
    \caption{Convergence of solvation free energies for phthalimide in (A) water and (B) octanol, comparing HREX, $\lambda$-OPES-Explore, and $\lambda$-OPES. Shaded regions represent the corresponding uncertainties.}
    \label{fig:w-biasing}
\end{figure}

Convergence profiles of the solvation free energies are highlighted in Figures~\ref{fig:w-biasing} A and B for phthalimide in water and in octanol, respectively, and are reported for all solutes in Figure~S2.
In water, both OPES protocols reach free energy plateaus consistent with HREX on similar timescales, indicating that either biasing strategy is sufficient for these rapidly relaxing aqueous systems.
In octanol, however, performance becomes more method-dependent.
For phthalimide, cinnamamide, and \textit{m}-methylphenol, $\lambda$-OPES-Explore converges more slowly and tends to overestimate the solvation free energy, while $\lambda$-OPES remains closer to HREX. For the other solutes, the differences are smaller and comparable to those observed in water.

\begin{figure}[!t]
    \centering
    \includegraphics[scale=0.85]{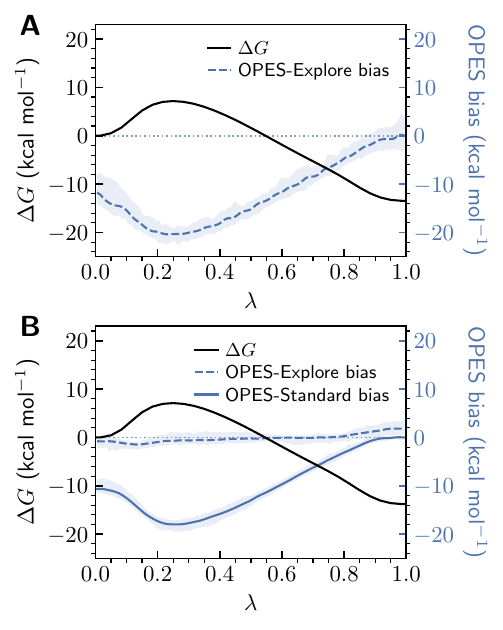}
    \caption{Cumulative free energy, $\Delta G$, and OPES bias potentials for phthalimide in octanol obtained from (A) $\lambda$-OPES-Explore and (B) $\lambda$-OPES simulations. Shaded regions indicate the time evolution of the bias potentials, and solid lines represent the final mean profiles.}
    \label{fig:energy-biasing}
\end{figure}

The stronger dependence on the biasing protocol in octanol is consistent with the slower relaxation and more persistent local structuring around the solute in this solvent. 
Under these conditions, accurate sampling requires not only exploration between the alchemical end states,
but also efficient equilibration throughout the alchemical pathway, including regions where the free energy profile is steeper, or the solvent response is more difficult to sample. To examine how the two OPES protocols achieve this sampling, we analyze the final free energy profile along $\lambda$ together with the corresponding OPES bias potentials for phthalimide in octanol. These results are shown in Figures~\ref{fig:energy-biasing}A and B for $\lambda$-OPES-Explore and $\lambda$-OPES, respectively; analogous analyses for all solutes are reported in Figure~S3.

\begin{figure}[!t]
    \centering
    \includegraphics[scale=0.85]{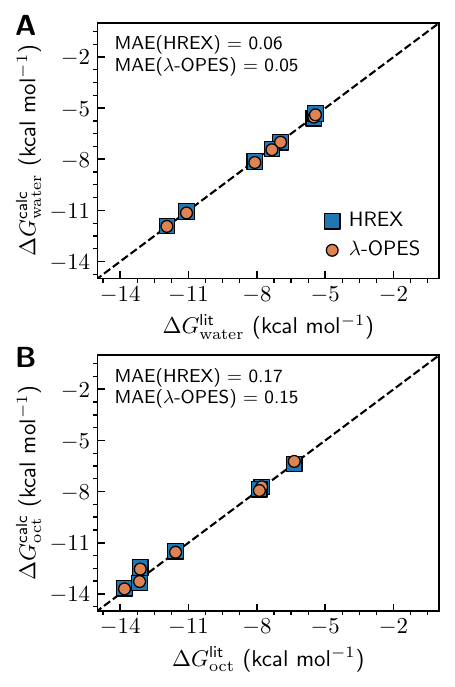}
    \caption{Calculated solvation free energies for the seven small organic solutes in (A) water and (B) octanol from HREX and $\lambda$-OPES simulations, compared with literature values.\cite{Bannan_2016} The dashed line indicates perfect agreement, and the mean absolute error (MAE) relative to the literature values is reported in each panel. Error bars are smaller than the symbol size and are omitted.}
    \label{fig:val-small}
\end{figure}

\begin{figure*}[!t]
    \centering
    \includegraphics[scale=0.85]{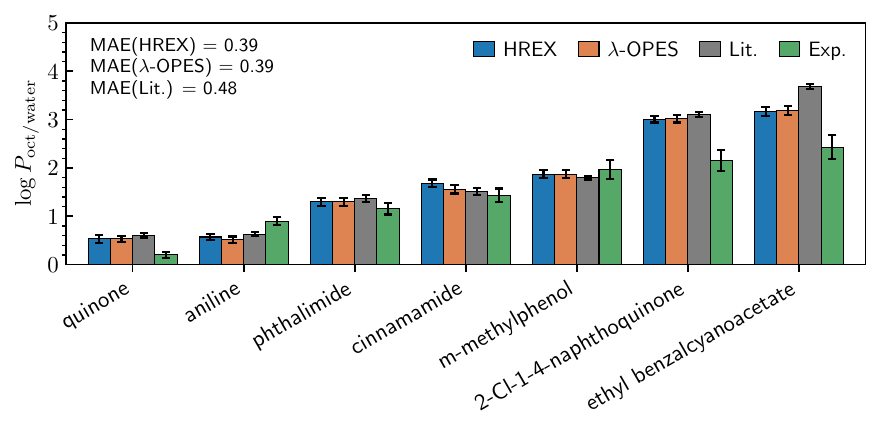}
    \caption{Octanol-water partition coefficients for the seven rigid organic solutes obtained from HREX and $\lambda$-OPES simulations, compared with literature \cite{Bannan_2016} and experimental \cite{Hansch_1995, Sangster_1989, Leo_1971} values. Mean absolute error (MAE) is reported relative to experiments. Error bars represent the uncertainties propagated from the solvation free energy calculations.}
    \label{fig:logP-small}
\end{figure*}

\begin{figure*}[!t]
    \centering
    \includegraphics[scale=0.85]{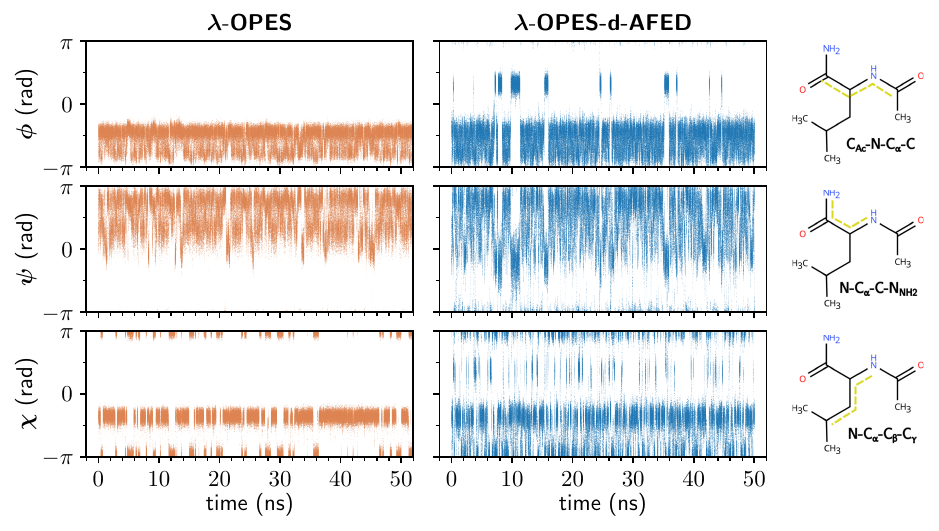}
    \caption{Time evolution of the $\phi$, $\psi$, and $\chi$ dihedral angles of Ac-Leu-NH$_2$ in octanol from $\lambda$-OPES and $\lambda$-OPES-d-AFED simulations. The highlighted backbone and side-chain dihedral angles are illustrated on the right.}
    \label{fig:dihedrals-peptides}
\end{figure*}

For $\lambda$-OPES-Explore, the exploratory bias exhibits a broader temporal dispersion and a noisier mean profile, indicating that OPES-Explore alone provides a less stable representation of the underlying free energy landscape. In the combined $\lambda$-OPES protocol, OPES-Standard provides the dominant contribution and exhibits an approximately opposite shape to the physical free energy profile, thereby reducing its variation along $\lambda$, while the weaker OPES-Explore component maintains exploration and promotes transitions between $\lambda=0$ and $\lambda=1$. This division of roles between free energy profile flattening and exploratory sampling explains the greater robustness of the combined protocol, particularly in octanol.

The OPES-based $\lambda$-dynamics protocols are at least as efficient as HREX for the systems considered here. Their main distinction lies instead in the sampling framework: HREX relies on multiple replicas at predefined alchemical states, whereas $\lambda$-dynamics samples a continuous alchemical coordinate within a single trajectory. The purpose of this comparison is to assess whether it can provide comparably reliable free energy estimates while avoiding the need to define and coordinate multiple discrete $\lambda$ windows. This single-trajectory formulation may be particularly advantageous when running many replicas concurrently is computationally demanding, for example for large or complex systems or when the potential energy evaluations rely on memory-intensive machine learning force fields. In such cases, the GPU memory and communication requirements associated with coordinating multiple HREX replicas may become a practical limitation.

We also examine the sensitivity of the dynamics to the mass of $\lambda$ using quinone in water as a controlled test case. Quinone is one of the smallest molecules of the set, and aqueous solvation is less affected by the slower solvent relaxation observed in octanol, allowing the effect of the $\lambda$ mass to be isolated more effectively. In addition to the value used above, 1 Da\,nm$^2$, two larger masses, $10^2$ and $10^3$ Da\,nm$^2$, are examined (Figure~S4). All three values yielded essentially the same converged solvation free energy, indicating that the equilibrium estimate is insensitive to the $\lambda$ mass within this range. 
As expected, increasing the $\lambda$ mass slows the extended-variable dynamics and reduces the frequency of transitions between the physical end states.

The same quinone-in-water system is used to evaluate $\lambda$-d-AFED under similarly controlled conditions. In this approach, $\lambda$ is coupled to a high auxiliary temperature to promote barrier crossing. Simulations are performed with $\lambda$ thermostatted at 5000 K (Figure~S5) and 10000 K (Figure~S6) and with masses of $10^4$, $10^5$, and $10^6$ Da\,nm$^2$. In contrast to $\lambda$-OPES, the calculated solvation free energies depend more sensitively on the $\lambda$ mass. Smaller masses led to overestimated free energies and less stable dynamics, indicating that $\lambda$-d-AFED, while as accurate as $\lambda$-OPES with a correct parameter choice, is more sensitive to the balance among mass selection, adiabatic separation, and high-temperature coupling. 

\subsection{Partition Coefficients of Small Organic Molecules}

Having examined the convergence of the different alchemical sampling strategies, we evaluate the final solvation free energies for the seven rigid organic solutes. The results obtained with HREX and $\lambda$-OPES are compared with the reference calculations of \citet{Bannan_2016} in Figures~\ref{fig:val-small} A and B and Table~S1. The two methods yield nearly indistinguishable free energies and reproduce the reference data with mean absolute deviations below 0.06 kcal mol$^{-1}$ in water and 0.17 kcal mol$^{-1}$ in octanol. The somewhat larger deviations observed in octanol are consistent with the greater sampling challenges associated with this solvent. The largest discrepancy, approximately 0.6 kcal mol$^{-1}$, occurs for ethyl benzalcyanoacetate, which is the largest solute in the set and contains rotatable bonds despite its predominantly rigid conjugated structure. Overall, the close agreement with the reference calculations indicates that the LBF alchemical model used here provides results consistent with those obtained using the soft-core protocol adopted by \citet{Bannan_2016}, despite the differences between their functional forms.

The resulting octanol-water partition coefficient values are shown in Figure~\ref{fig:logP-small} and Table~S2. Because HREX and $\lambda$-OPES yield very similar solvation free energies, the corresponding log$P_{\mathrm{oct/water}}$ are also nearly identical and remain close to the literature estimates. Comparison with experimental data\cite{Hansch_1995, Sangster_1989, Leo_1971} shows a systematic overestimation of log$P_{\mathrm{oct/water}}$ by the GAFF + TIP3P model. This trend suggests that the remaining discrepancies arise mainly from the force field description of molecular interactions rather than from incomplete sampling along $\lambda$. Nevertheless, the simulations reproduce the qualitative ordering of hydrophobicity across the series, indicating that the present protocol captures the dominant thermodynamic trends governing the transfer between water and octanol.

\subsection{Conformational Effects in Acetyl Amino-Acid Amides}

For the rigid solutes discussed above, accurate solvation free energies primarily require converged sampling along the alchemical coordinate. In acetyl amino-acid amides, however, backbone and side-chain flexibility introduce additional conformational contributions, since different conformers can be stabilized differently in water and octanol. Incomplete sampling of these conformational states may affect both the individual solvation free energies and the resulting octanol-water partition coefficients. To assess this effect, we compare the previously established $\lambda$-OPES protocol with $\lambda$-OPES-d-AFED, in which $\lambda$-OPES is retained for alchemical sampling while d-AFED is applied to selected dihedral angles.

Seven acetyl amino-acid amides are simulated in water and octanol using both protocols. For Ac-Ala-NH$_2$ and Ac-Gly-NH$_2$, d-AFED is applied to the backbone dihedrals $\phi$ (C$_{\mathrm{Ac}}$--N--C$_\alpha$--C) and $\psi$ (N--C$_\alpha$--C--N$_{\mathrm{NH_2}}$). For the remaining systems, the side-chain dihedral $\chi$ (N--C$_\alpha$--X--Y) is also included.
The effect of d-AFED on conformational sampling is illustrated for Ac-Leu-NH$_2$ in octanol in Figure~\ref{fig:dihedrals-peptides}, which compares the time evolution of $\phi$, $\psi$, and $\chi$. With $\lambda$-OPES alone, the trajectories remain confined to more localized angular regions. Adding d-AFED leads to broader exploration of the dihedral space and more frequent transitions between conformational states. The same trend is observed for the full set of acetyl amino-acid amides in water and octanol, as reported in Figures~S7 and S8, respectively.

\begin{figure}[!t]
    \centering
    \includegraphics[scale=0.85]{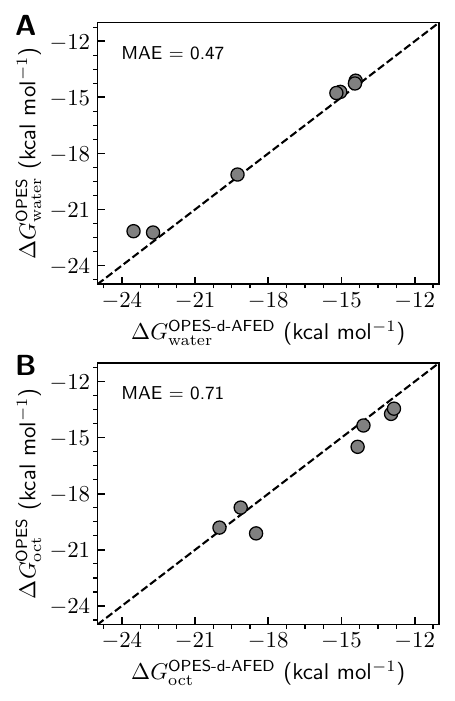}
    \caption{Comparison of solvation free energies obtained from $\lambda$-OPES and $\lambda$-OPES-d-AFED simulations for acetyl amino-acid amides in (A) water and (B) octanol. The dashed line indicates perfect agreement, and the mean absolute error (MAE) between the two protocols is reported in each panel. Error bars are smaller than the symbol size and are omitted.}
    \label{fig:deltaG-peptides}
\end{figure}

The improved exploration of conformational space leads to systematic changes in the calculated solvation free energies. Although both protocols reach stable plateaus within approximately 30--50 ns (Figure~S9), the converged values are not equivalent. Figures~\ref{fig:deltaG-peptides} A and B compare the solvation free energies obtained with $\lambda$-OPES and $\lambda$-OPES-d-AFED in water and octanol, respectively. The mean absolute deviations between the two protocols are 0.47 kcal mol$^{-1}$ in water and 0.71 kcal mol$^{-1}$ in octanol, indicating that conformational sampling has a larger effect in octanol. The $\lambda$-dependent free energy profiles and OPES bias potentials (Figure~S10) remain consistent with the behavior observed for the rigid organic solutes, with OPES-Standard providing the dominant bias along $\lambda$ and OPES-Explore maintaining exploratory sampling.

\begin{figure*}[!t]
    \centering
    \includegraphics[scale=0.85]{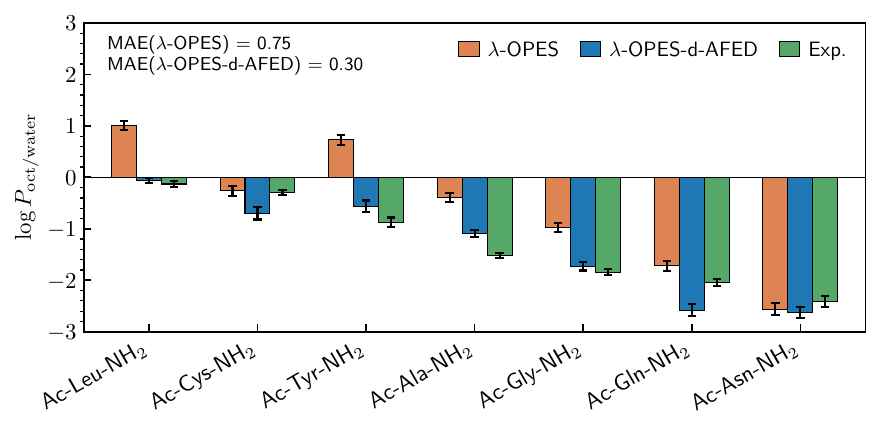}
    \caption{Octanol-water partition coefficients for acetyl amino-acid amides obtained from $\lambda$-OPES and $\lambda$-OPES-d-AFED simulations, compared with experimental values.\cite{Fauchere_1983} Mean absolute errors are reported relative to experiment, and error bars represent uncertainties propagated from the solvation free energy calculations.}
    \label{fig:logP-peptides}
\end{figure*}

The impact of conformational sampling becomes even more apparent in the octanol-water partition coefficients, which are compared with experimental values\cite{Fauchere_1983} in Figure~\ref{fig:logP-peptides}. Including d-AFED systematically lowers the predicted log$P_{\mathrm{oct/water}}$ values and improves agreement with experiment. This correction is particularly important for Ac-Leu-NH$_2$ and Ac-Tyr-NH$_2$: $\lambda$-OPES predicts positive partition coefficients for these compounds, whereas $\lambda$-OPES-d-AFED shifts them to negative values, consistent with the experimental trend. As a result, the mean absolute error decreases from 0.75 with $\lambda$-OPES to 0.30 with $\lambda$-OPES-d-AFED.
Despite this overall improvement, the partition coefficients of Ac-Cys-NH$_2$ and Ac-Gln-NH$_2$ remain overestimated, indicating some remaining solute-specific deviations.
These results demonstrate that reliable partition coefficients for flexible peptide-like molecules require both converged alchemical sampling and adequate exploration of the relevant conformational space. They also indicate that when combined with enhanced conformational sampling, the AMBER14SB/OPC/GAFF2 model provides an effective description of the acetyl amino acid amide series in water and octanol.

\section{Conclusions}

We have developed an integrated enhanced-sampling framework for solvation free energy calculations in which solvent mass scaling, OPES-biased $\lambda$-dynamics, and d-AFED-based conformational acceleration are combined within a unified protocol. The method was evaluated in water and octanol for systems ranging from rigid organic solutes to flexible acetyl amino acid amides, allowing the separate roles of solvent relaxation, alchemical sampling, and solute conformational sampling to be assessed.

For octanol, reducing the solvent atomic masses by a factor of ten substantially accelerated convergence in HREX simulations, decreasing the simulation time required to reach stable solvation free energies by more than fivefold. The final free energy estimates remained unchanged within uncertainty, confirming that the improvement is dynamical rather than thermodynamic. This result indicates that mass scaling provides a simple and general strategy for accelerating relaxation in viscous organic solvents without changing the target equilibrium configurational distribution.

We then developed and tested OPES-biased $\lambda$-dynamics protocols in which the alchemical coordinate is propagated as a dynamical variable. The single-bias $\lambda$-OPES-Explore protocol promoted transitions along $\lambda$ but was less robust in octanol, where slow solvent relaxation made convergence more sensitive to the biasing strategy. The best performance was obtained with the combined $\lambda$-OPES protocol, in which OPES-Standard provides the dominant contribution to the adaptive bias, enabling rapid and stable convergence of the free energy profile, while the weaker OPES-Explore component primarily increases transitions between the alchemical end states. For the rigid organic solutes, this protocol produced solvation free energies and octanol-water partition coefficients in close agreement with HREX reference calculations, while avoiding predefined $\lambda$ windows and multiple parallel replicas. The systematic overestimation of experimental log$P_{\mathrm{oct/water}}$ values for these compounds is likely dominated by force field limitations rather than by incomplete alchemical sampling.

For flexible acetyl amino-acid amides, coupling $\lambda$-OPES with d-AFED on selected backbone and side-chain dihedrals led to broader conformational exploration and systematic changes in the calculated solvation free energies. The effect was larger in octanol than in water, highlighting the stronger coupling between conformational relaxation and solvation in the more viscous and heterogeneous organic phase. These conformational corrections had a direct impact on the predicted partition coefficients: $\lambda$-OPES-d-AFED reduced the mean absolute error relative to experiment from 0.75 to 0.30 log units and correctly shifted the predicted partitioning of Ac-Leu-NH$_2$ and Ac-Tyr-NH$_2$ toward the experimentally observed water-preferred regime.

These results show that accurate solvation and partition free energies for flexible molecules require simultaneous treatment of alchemical and conformational sampling. The proposed $\lambda$-OPES-d-AFED framework provides a practical route to achieve this goal within a single extended-dynamics simulation, while solvent mass scaling further improves efficiency in slowly relaxing organic phases. More broadly, the method provides a foundation for applying alchemical free energy calculations to larger and more conformationally complex solutes, in which solvent relaxation, intramolecular barriers, and alchemical transitions are strongly coupled.

\section*{Author contributions}
\textbf{Gabriela B. Correa}: Data curation; Formal analysis; Investigation; Methodology; Software; Writing – original draft.
\textbf{Charlles R. A. Abreu}: Conceptualization; Software; Supervision; Writing – review \& editing.
\textbf{Nisanth N. Nair}: Conceptualization; Supervision; Writing – review \& editing.
\textbf{Mark E. Tuckerman}: Conceptualization; Funding acquisition; Project administration; Supervision; Writing – review \& editing.

\section*{Acknowledgments}
GBC and MET were funded by Breakthrough Electrolytes for Energy Storage Systems—an Energy Frontier Research Center of the United States Department of Energy, Office of Science, Basic Energy Sciences under Award \#DE-SC0019409.

\section*{Supplementary Material}
Convergence of solvation free energies for rigid organic molecules, including the effects of octanol mass scaling, $\lambda$ mass, and auxiliary temperature; free energy and OPES bias profiles; solvation free energies and octanol--water partition coefficients; dihedral sampling of $N$-acetyl amino-acid amides; and comparison of $\lambda$-OPES and $\lambda$-OPES-d-AFED results.

\section*{Conflicts of interest}
The authors have no conflicts to disclose.

\section*{Data availability}
The data supporting this article have been included as part of the Supplementary Material.
The simulation scripts are available at \url{github.com/gabriela-correa/lambda-OPES-d-AFED}.

\bibliography{bib.bib}

\end{document}


\title{
\textbf{Supplementary Information} \\[1em]
Integrated Alchemical and Conformational Enhanced Sampling for Solvation Free Energy Calculations
}

\author{Gabriela B. Correa}
\affiliation{Department of Chemistry, New York University, New York City, New York 10003, United States}

\author{Charlles R. A. Abreu}
\affiliation{Department of Chemical Engineering, Massachusetts Institute of Technology, Cambridge, Massachusetts 02139, United States}

\author{Nisanth N. Nair}
\affiliation{Department of Chemistry, Indian Institute of Technology Kanpur, Kanpur 208016, India}

\author{Mark E. Tuckerman}
\email{mark.tuckerman@nyu.edu}
\affiliation{Department of Chemistry, New York University, New York City, New York 10003, United States}
\affiliation{Simons Center for Computational Physical Chemistry, New York University, New York City, New York 10003, United States}
\affiliation{Courant Institute of Mathematical Sciences, New York University, New York City, New York 10012, United States}

\maketitle

\clearpage

\begin{figure}[H]
    \centering
    \includegraphics[scale=0.95]{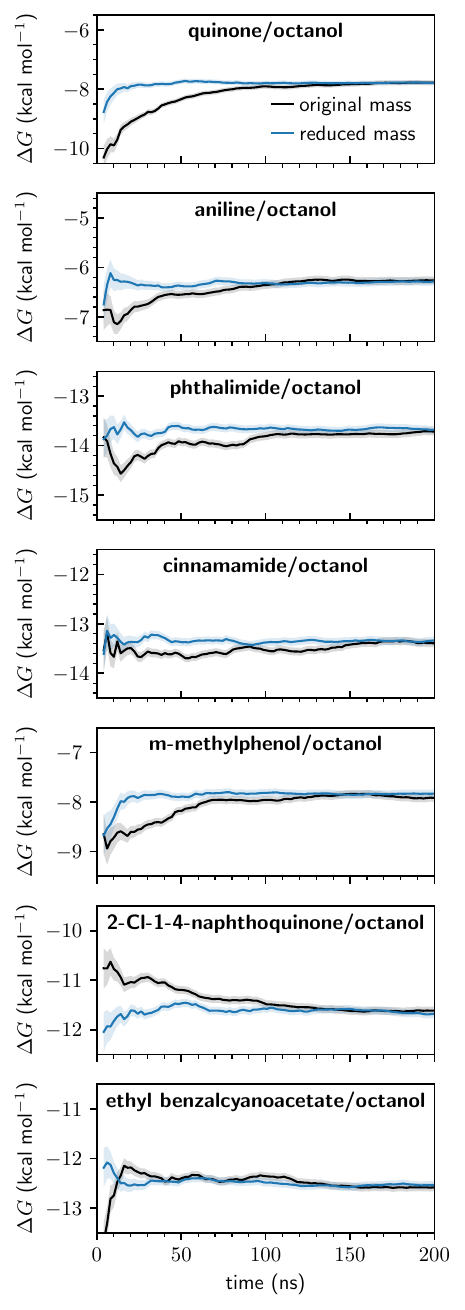}
    \caption{Convergence of solvation free energies in octanol for small organic molecules from HREX simulations, comparing the original octanol mass and a 10-fold reduced octanol mass. The shaded regions represent the uncertainties.}
    \label{fig:scaling-si}
\end{figure}

\begin{figure}[H]
    \centering
    \includegraphics[scale=0.95]{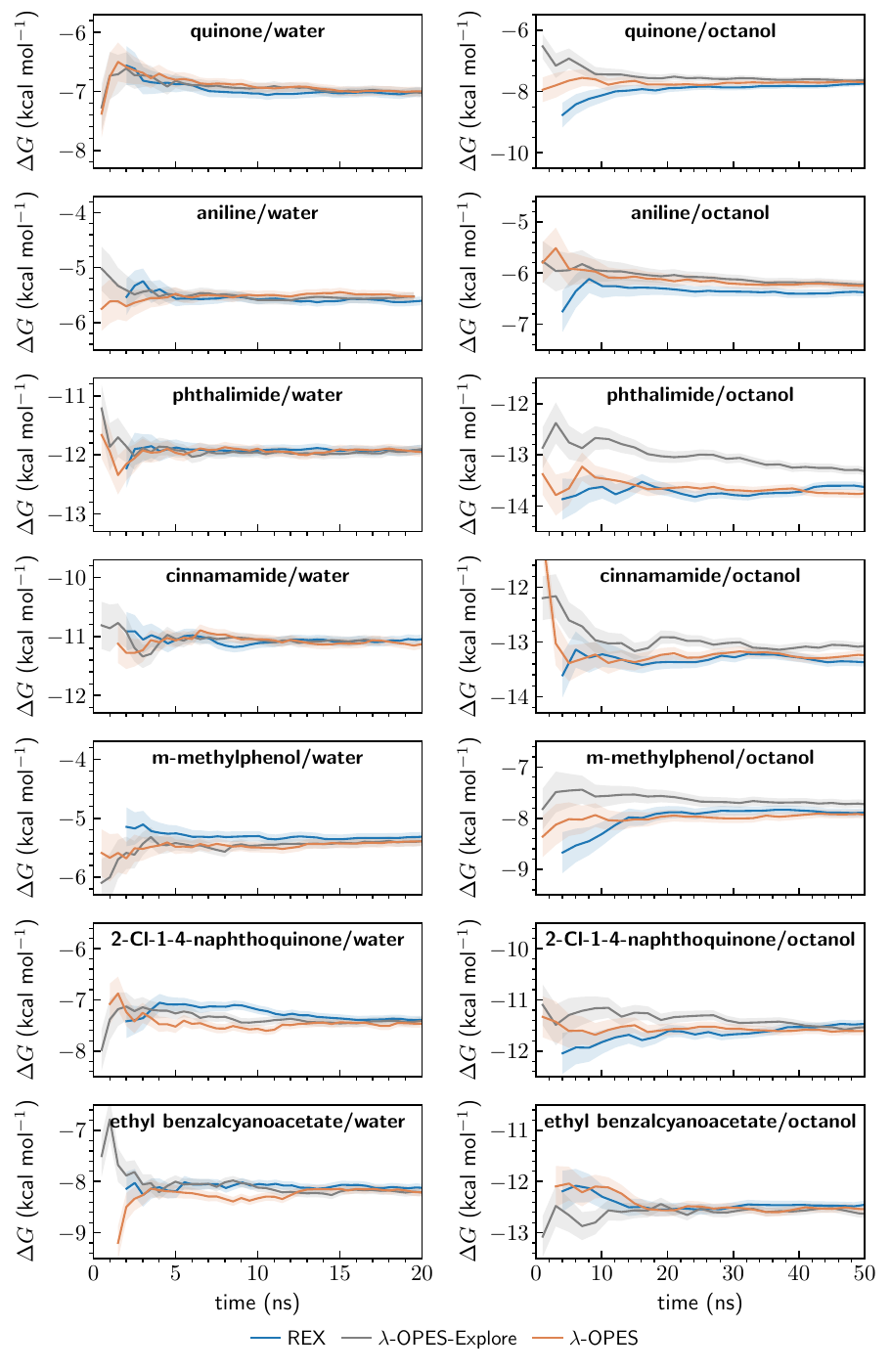}
    \caption{Convergence of solvation free energies in water and octanol for small organic molecules, comparing HREX, $\lambda$-OPES-Explore, and $\lambda$-OPES. The shaded regions represent the uncertainties.}
    \label{fig:al-cost}
\end{figure}

\clearpage

\begin{figure}[htb!]
    \centering
    \includegraphics[scale=0.95]{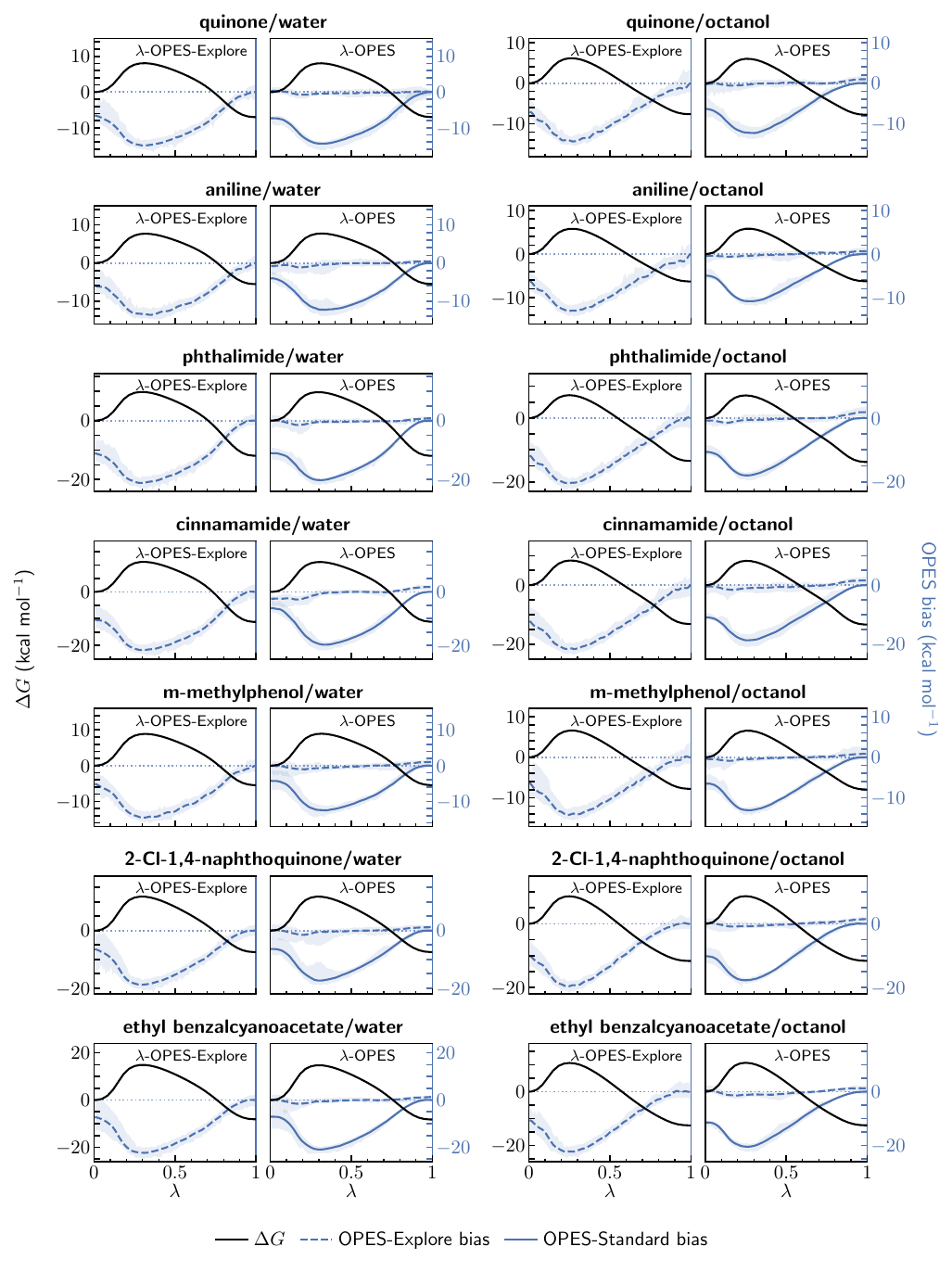}
    \caption{Free energy profiles along $\lambda$ and corresponding OPES bias potentials for small organic molecules in water and octanol. The left panels show $\lambda$-OPES-Explore and the right panels show the combined $\lambda$-OPES protocol. Shaded regions indicate the time evolution of the bias potentials, and solid lines represent the final mean profiles.}
    \label{fig:al-cost}
\end{figure}

\begin{figure}[htb!]
    \centering
    \includegraphics[scale=0.95]{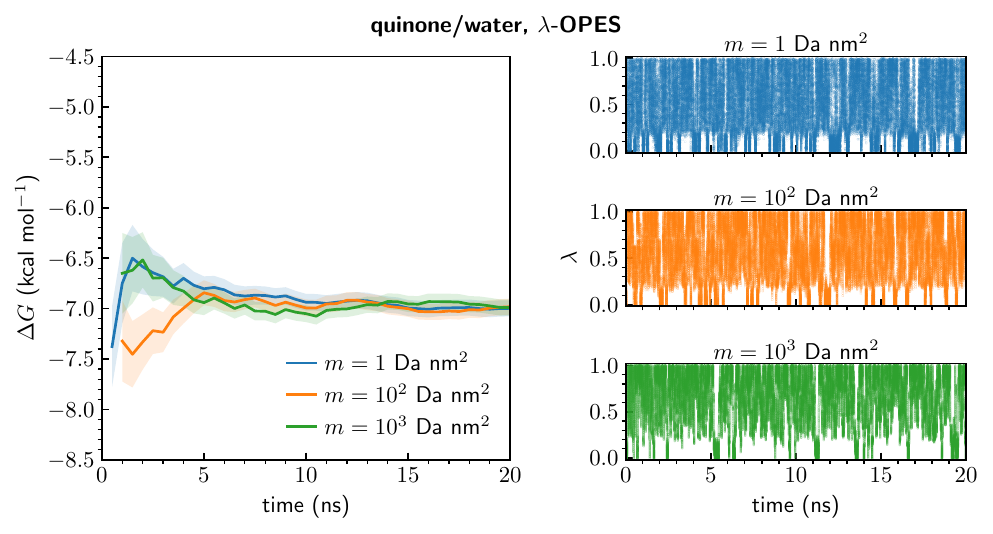}
    \caption{Solvation free energy and $\lambda$ trajectories for quinone in water obtained from $\lambda$-OPES calculations with $\lambda$ masses of 1, $10^2$, and $10^3$ Da\,nm$^2$. The shaded regions represent the uncertainties.}
    \label{fig:al-cost}
\end{figure}

\begin{figure}[htb!]
    \centering
    \includegraphics[scale=0.95]{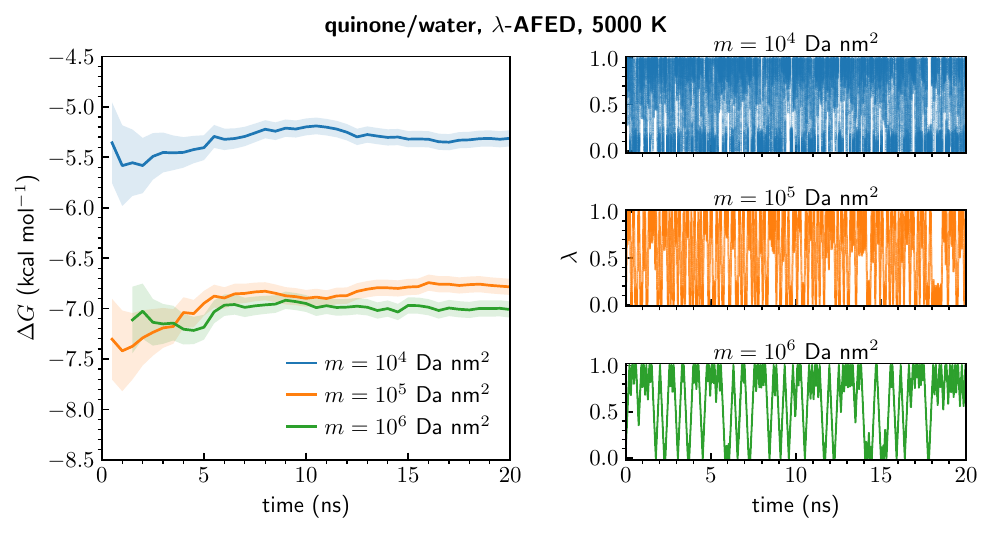}
    \caption{Solvation free energy and $\lambda$ trajectories for quinone in water obtained from $\lambda$-AFED calculations at 5000 K with $\lambda$ masses of $10^4$, $10^5$, and $10^6$ Da\,nm$^2$. The shaded regions represent the uncertainties.}
    \label{fig:al-cost}
\end{figure}

\begin{figure}[htb!]
    \centering
    \includegraphics[scale=0.95]{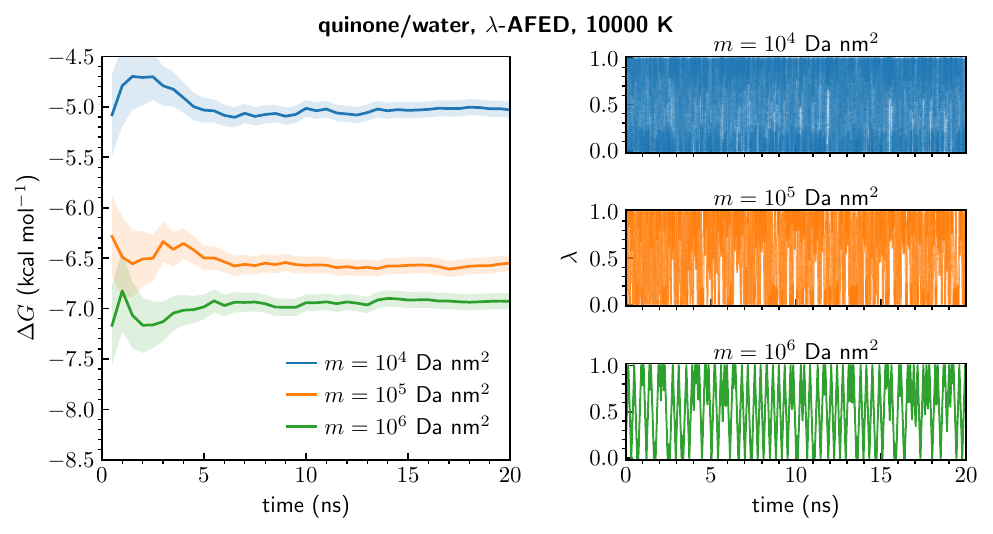}
    \caption{Solvation free energy and $\lambda$ trajectories for quinone in water obtained from $\lambda$-AFED calculations at 10000 K with $\lambda$ masses of $10^4$, $10^5$, and $10^6$ Da\,nm$^2$. The shaded regions represent the uncertainties.}
    \label{fig:al-cost}
\end{figure}

\clearpage

\begin{table}[h!]
\small
\centering
\caption{
Final solvation free energies in water and octanol for small organic molecules obtained using the $\lambda$-OPES and HREX methods; corresponding values from the MD literature are also reported.
}
\label{tab:small_free_energy}
\begin{tabular*}{\linewidth}{@{\extracolsep{\fill}}lcccccc}
\hline
 & \multicolumn{3}{c}{$\Delta G_\text{water}$ (kcal mol$^{-1}$)} 
 & \multicolumn{3}{c}{$\Delta G_\text{oct}$ (kcal mol$^{-1}$)} \\
\cline{2-4}
\cline{5-7}
Solute 
 & $\lambda$-OPES & HREX & Lit.\cite{Bannan_2016} 
 & $\lambda$-OPES & HREX & Lit.\cite{Bannan_2016}   \\
\hline
quinone                        & $-7.00_5$  & $-7.01_{6}$   & $-6.96_{2}$  & $-7.72_{8}$   & $-7.74_{10}$   & $-7.79_{7}$  \\
aniline                        & $-5.53_6$  & $-5.60_{6}$   & $-5.50_{2}$  & $-6.23_8$     & $-6.38_{8}$    & $-6.36_{5}$  \\
phthalimide                    & $-11.94_5$ & $-11.91_{5}$  & $-11.95_{2}$ & $-13.71_{10}$ & $-13.68_{10}$  & $-13.81_{10}$ \\
cinnamamide                    & $-11.15_6$ & $-11.07_{5}$  & $-11.09_{2}$ & $-13.28_{10}$ & $-13.36_{10}$  & $-13.15_{10}$ \\
m-methylphenol                 & $-5.39_5$  & $-5.32_{5}$   & $-5.43_{2}$  & $-7.94_{10}$  & $-7.87_{10}$   & $-7.89_{5}$  \\
2-Cl-1-4-napthoquinone         & $-7.44_6$  & $-7.39_{6}$   & $-7.34_{2}$  & $-11.56_{9}$  & $-11.50_{8}$   & $-11.57_{7}$ \\
ethyl benzalcyanoacetate       & $-8.19_6$  & $-8.13_{6}$   & $-8.09_{2}$  & $-12.55_{10}$ & $-12.46_{10}$  & $-13.12_{7}$  \\
\hline
\end{tabular*}
\end{table}

\begin{table*}[h!]
\small
\centering
\caption{
log$P_{\mathrm{oct/water}}$ for small organic molecules obtained using the $\lambda$-OPES and HREX methods; corresponding values from the MD literature and experiments are also reported.
}
\label{tab:small_partition}
\begin{tabular*}{\linewidth}{@{\extracolsep{\fill}}lcccc}
\hline
Solute
 & $\lambda$-OPES 
 & HREX
 & Lit.\cite{Bannan_2016}
 & Exp.\cite{Hansch_1995, Sangster_1989, Leo_1971} \\
\hline
quinone                        &   $0.53_7$  & $0.53_9$    & $0.61_5$ & 0.20 \\
aniline                        &   $0.51_7$ & $0.57_7$    & $0.63_4$ & 0.90 \\
phthalimide                    &   $1.30_8$ & $1.30_8$    & $1.36_7$ & 1.15 \\
cinnamamide                    &   $1.56_9$ & $1.68_8$    & $1.52_7$ & 1.43 \\
m-methylphenol                 &   $1.87_8$ & $1.87_8$    & $1.80_4$ & 1.96 \\
2-Cl-1-4-napthoquinone         &   $3.02_8$ & $3.01_7$    & $3.10_5$ & 2.15 \\
ethyl benzalcyanoacetate       &   $3.19_9$ & $3.17_8$    & $3.69_5$ & 2.43 \\
\hline
\end{tabular*}
\end{table*}

\clearpage

\begin{figure}[htb!]
    \centering
    \includegraphics[scale=0.95]{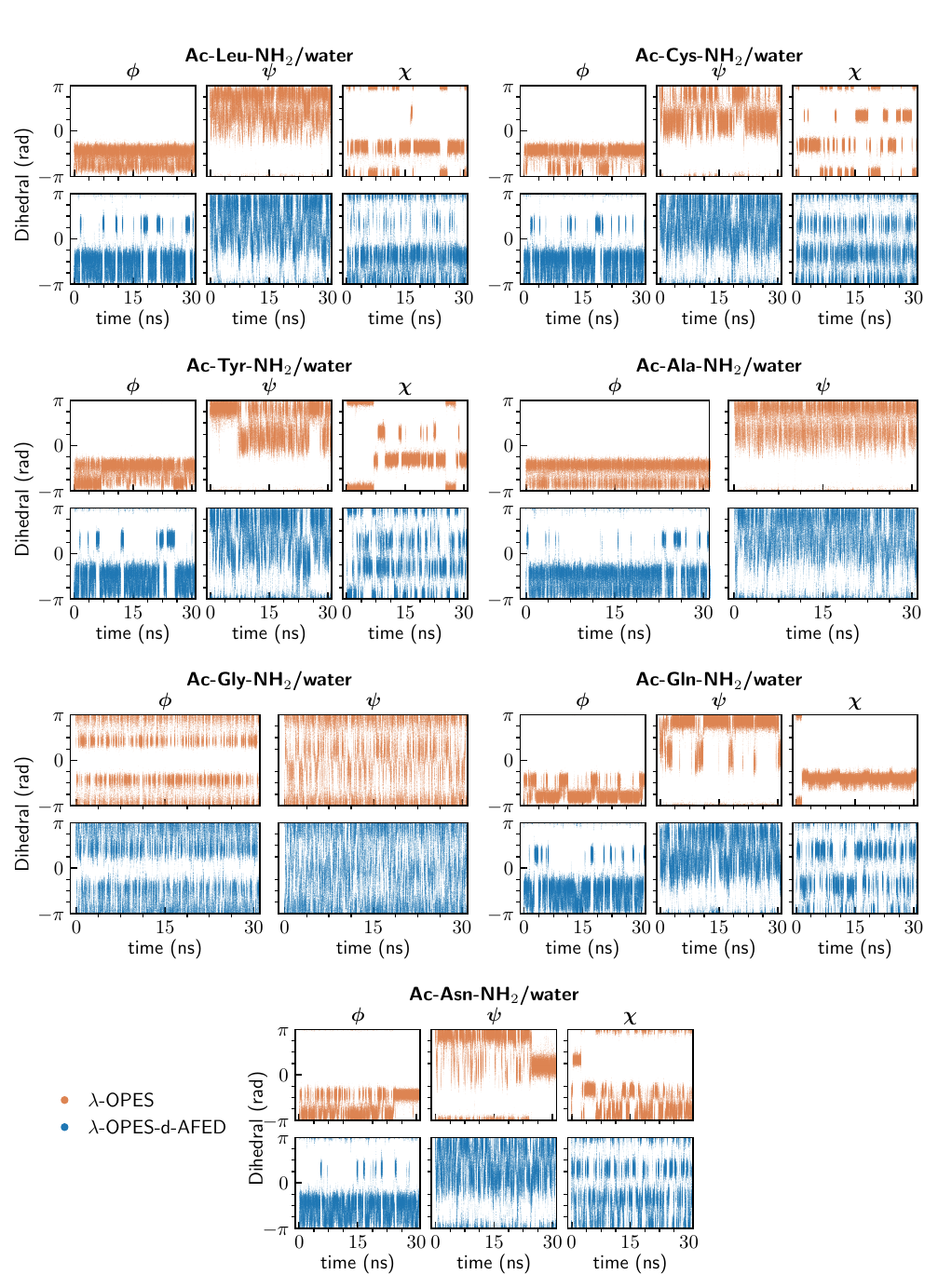}
    \caption{Time evolution of the $\phi$, $\psi$, and $\chi$ dihedral angles of $N$-acetyl amino-acid amides in water from $\lambda$-OPES and $\lambda$-OPES-d-AFED simulations.}
    \label{fig:al-cost}
\end{figure}

\begin{figure}[htb!]
    \centering
    \includegraphics[scale=0.95]{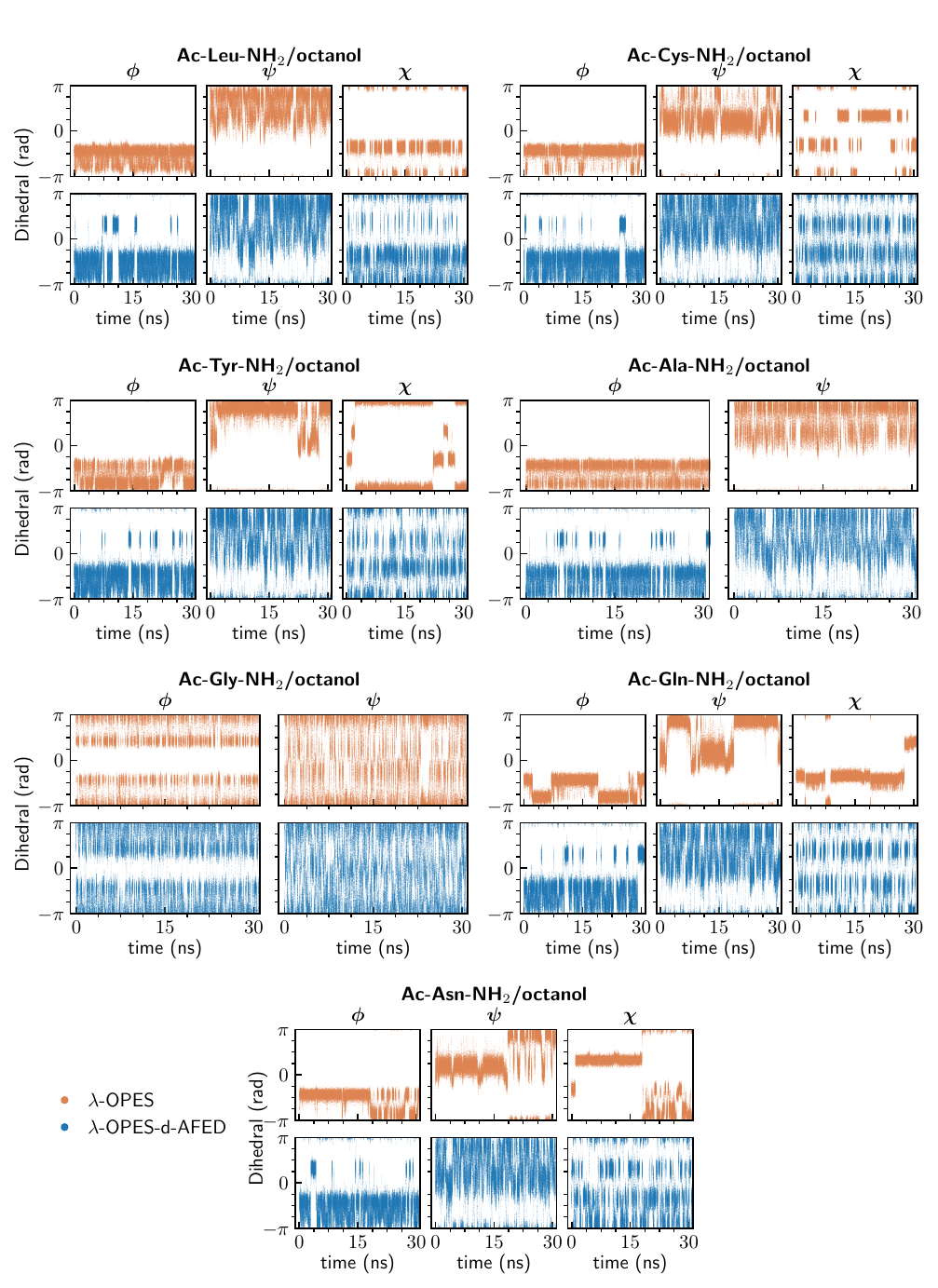}
    \caption{Time evolution of the $\phi$, $\psi$, and $\chi$ dihedral angles of $N$-acetyl amino-acid amides in octanol from $\lambda$-OPES and $\lambda$-OPES-d-AFED simulations.}
    \label{fig:al-cost}
\end{figure}

\begin{figure}[htb!]
    \centering
    \includegraphics[scale=0.95]{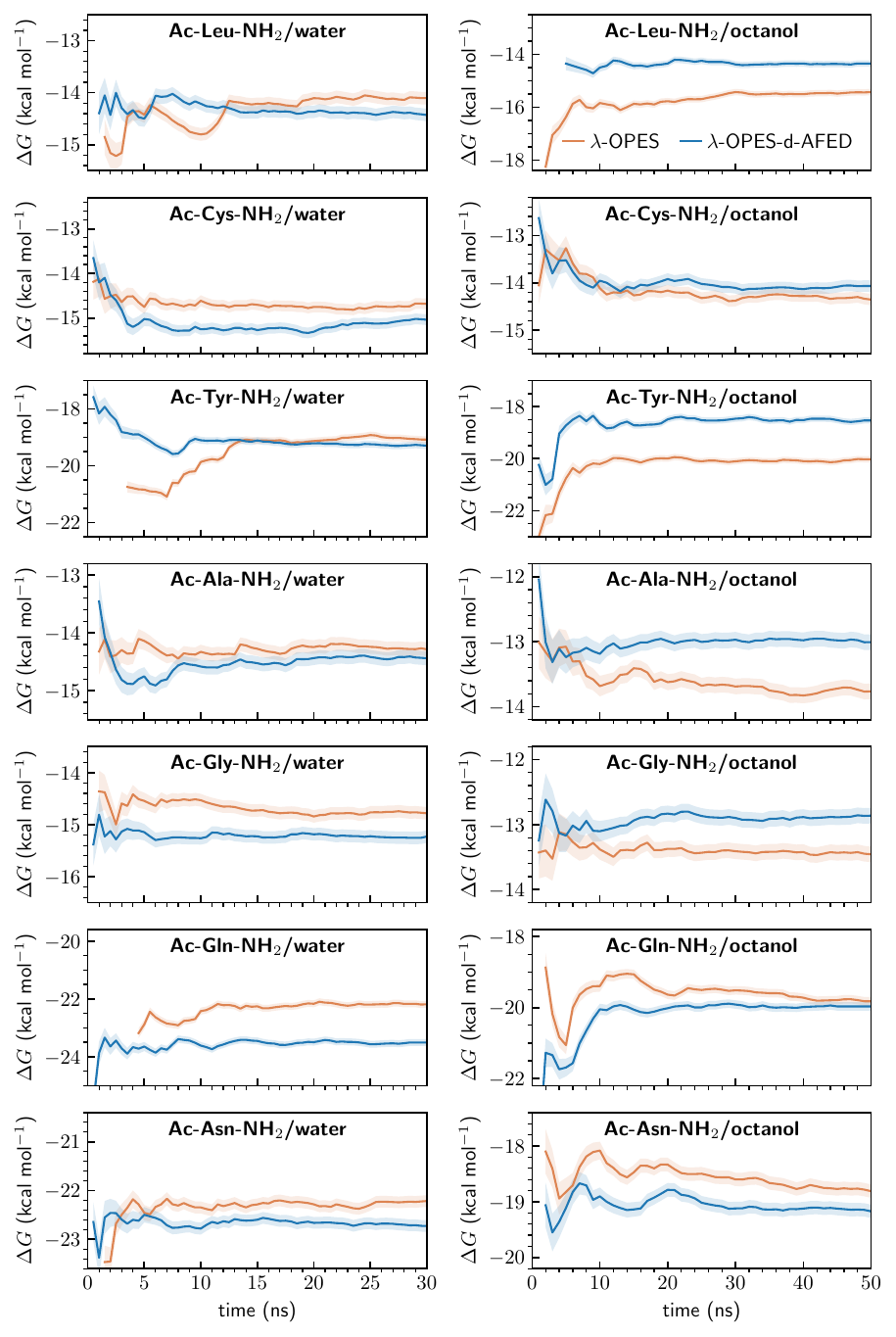}
    \caption{Convergence of solvation free energies in water and octanol for $N$-acetyl amino-acid amides, comparing $\lambda$-OPES and $\lambda$-OPES-d-AFED. The shaded regions represent the uncertainties.}
    \label{fig:al-cost}
\end{figure}

\begin{figure}[htb!]
    \centering
    \includegraphics[scale=0.95]{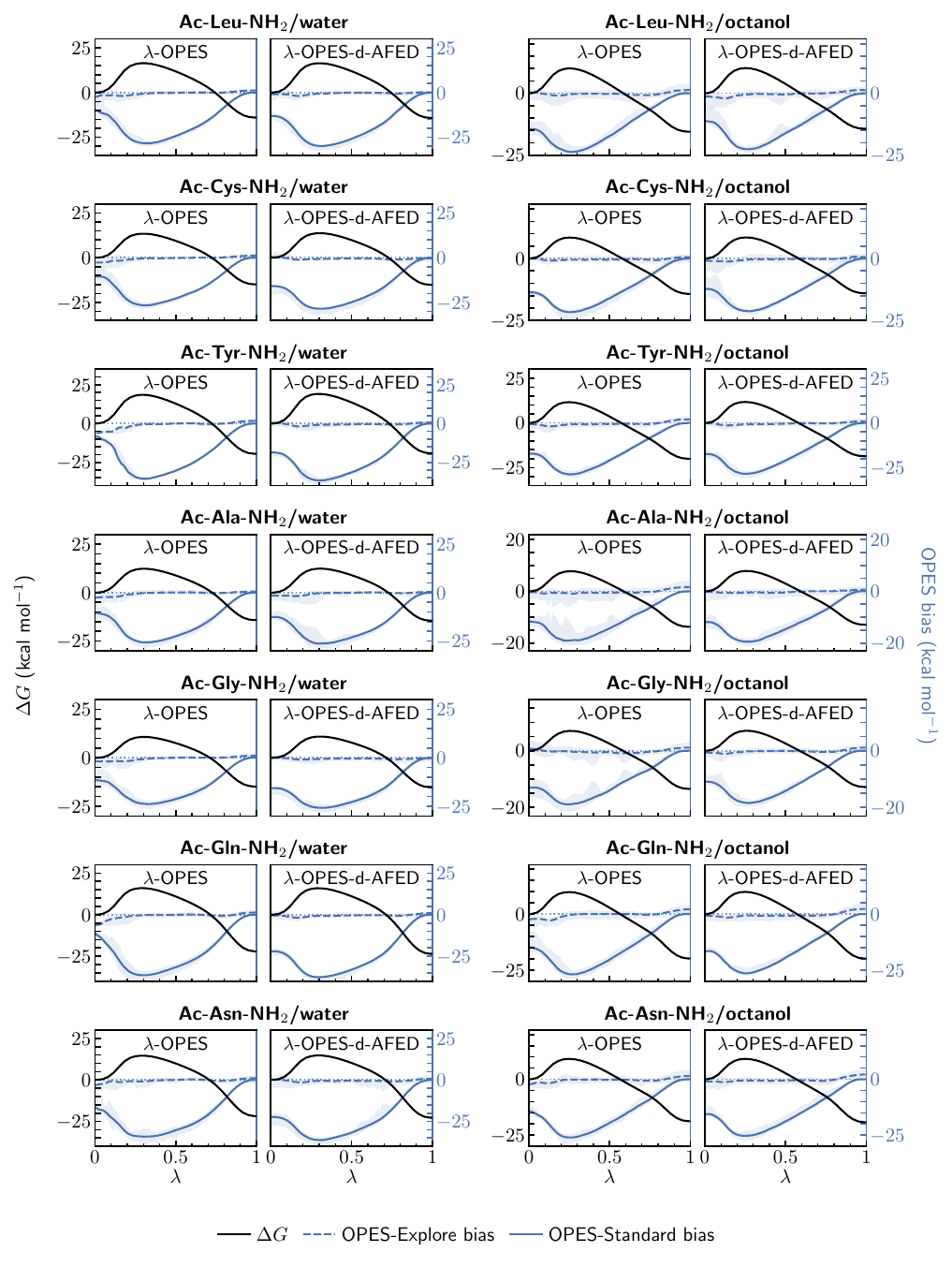}
    \caption{Free energy profiles along $\lambda$ and corresponding OPES bias potentials for the acetyl amino-acid amides in water and octanol. The left panels show $\lambda$-OPES and the right panels show $\lambda$-OPES-d-AFED results. Shaded regions indicate the time evolution of the bias potentials, and solid lines represent the final mean profiles.}
    \label{fig:al-cost}
\end{figure}

\clearpage

\begin{table}[h!]
\small
\centering
\caption{
Final solvation free energies in water and octanol for acetyl amino-acid amides obtained using the $\lambda$-OPES and $\lambda$-OPES-d-AFED methods.
}
\label{tab:peptides_free_energy}
\begin{tabular*}{\linewidth}{@{\extracolsep{\fill}}lcccc}
\hline
 & \multicolumn{2}{c}{$\Delta G_\text{water}$ (kcal mol$^{-1}$)} 
 & \multicolumn{2}{c}{$\Delta G_\text{oct}$ (kcal mol$^{-1}$)} \\
\cline{2-3}
\cline{4-5}
Solute 
 & $\lambda$-OPES & $\lambda$-OPES-d-AFED
 & $\lambda$-OPES & $\lambda$-OPES-d-AFED   \\
\hline
Acetyl-leucine amide        &  $-14.11_{9}$   &  $-14.42_5$      & $-15.49_9$    & $-14.34_{5}$      \\
Acetyl-cysteine amide       &  $-14.71_{9}$   &  $-15.05_9$      & $-14.35_{11}$ & $-14.10_{14}$      \\
Acetyl-tyrosine amide       &  $-19.13_{9}$   &  $-19.26_{10}$   & $-20.12_{11}$ & $-18.50_{12}$    \\
Acetyl-alanine amide        & $-14.26_{8}$  &  $-14.45_{8}$    & $-13.73_{9}$  & $-12.97_{6}$     \\
Acetyl-glycine amide        &  $-14.77_{9}$   &  $-15.21_{6}$    & $-13.45_{8}$  & $-12.85_{10}$      \\ 
Acetyl-glutamine amide      &  $-22.16_{7}$   &  $-23.52_{13}$   & $-19.81_{12}$ & $-20.00_{9}$      \\
Acetyl-aspargine amide      &  $-22.23_{11}$   &  $-22.72_{11}$    & $-18.74_{12}$ & $-19.13_{11}$     \\
\hline
\end{tabular*}
\end{table}

\begin{table}[h!]
\small
\centering
\caption{
log$P_{\mathrm{oct/water}}$ for acetyl amino-acid amides obtained using the $\lambda$-OPES and $\lambda$-OPES-d-AFED methods; corresponding values from experiments are also reported.
}
\label{tab:peptides_partition}
\begin{tabular*}{\linewidth}{@{\extracolsep{\fill}}lccc}
\hline
Solute
 & $\lambda$-OPES  
 & $\lambda$-OPES-d-AFED
 & Exp.\cite{Fauchere_1983} \\
\hline
Acetyl-leucine amide    &  $1.01_9$  & $-0.06_{5}$ & $-0.13_2$  \\
Acetyl-cysteine amide   &  $-0.26_{10}$  & $-0.70_{12}$ & $-0.29_5$  \\
Acetyl-tyrosine amide   &  $0.73_{10}$   & $-0.56_{11}$  & $-0.87_9$  \\
Acetyl-alanine amide    & $-0.39_9$   & $-1.09_7$ & $-1.52_2$  \\
Acetyl-glycine amide    & $-0.97_9$   & $-1.73_8$ & $-1.84_2$  \\
Acetyl-glutamine amide  & $-1.72_{10}$   & $-2.58_{12}$ & $-2.05_7$  \\
Acetyl-aspargine amide  & $-2.56_{12}$   & $-2.63_{11}$ & $-2.41_{11}$  \\
\hline
\end{tabular*}
\end{table}

\clearpage

\bibliography{bib.bib}